\documentclass[onefignum,onetabnum]{siamart171218}

\usepackage{amsmath,amsfonts,amssymb}
\usepackage{textgreek}
\usepackage[frozencache=true,cachedir=minted-cache]{minted}

\let\savebigtimes\bigtimes

\let\bigtimes\relax

\usepackage{mathabx}

\let\bigtimes\savebigtimes

\usepackage{bm}
\usepackage{url}

\usepackage{algpseudocode}

\makeatletter
\AtBeginEnvironment{noerr}{\dontdofcolorbox}
\def\dontdofcolorbox{\renewcommand\fcolorbox[4][]{##4}}
\makeatother

\newenvironment{noerr}{}

\usepackage{lipsum}
\usepackage{amsfonts}
\usepackage{graphicx}
\usepackage{epstopdf}
\ifpdf
  \DeclareGraphicsExtensions{.eps,.pdf,.png,.jpg}
\else
  \DeclareGraphicsExtensions{.eps}
\fi

\newsiamremark{remark}{Remark}
\newsiamremark{hypothesis}{Hypothesis}
\crefname{hypothesis}{Hypothesis}{Hypotheses}
\newsiamthm{claim}{Claim}

\headers{NFFT.\MakeLowercase{jl}: Generic and Fast Implementation of the NFFT}{T. Knopp, M. Boberg, M, Grosser}

\title{NFFT.\MakeLowercase{jl}: Generic and Fast Julia Implementation of the  Nonequidistant Fast Fourier Transform}

\author{Tobias Knopp\thanks{University Medical Center Hamburg-Eppendorf and Hamburg University of Technology, Hamburg, Germany 
  (\email{t.knopp@uke.de}, \url{http://www.tuhh.de/ibi}).} 
\and Marija Boberg$^*$ \and Mirco Grosser$^*$
  }

\usepackage{amsopn}

\makeatletter
\newcommand*{\addFileDependency}[1]{
  \typeout{(#1)}
  \@addtofilelist{#1}
  \IfFileExists{#1}{}{\typeout{No file #1.}}
}
\makeatother

\newcommand*{\myexternaldocument}[1]{%
    \externaldocument{#1}%
    \addFileDependency{#1.tex}%
    \addFileDependency{#1.aux}%
}

\ifpdf
\hypersetup{
  pdftitle={NFFT.jl: Generic and Fast Julia Implementation of the  Nonequidistant Fast Fourier Transform},
  pdfauthor={T. Knopp}
}
\fi

\myexternaldocument{ek_supplement}

\begin{document}

\maketitle

\begin{abstract}
  The non-equidistant fast Fourier transform (NFFT) is an extension of the famous fast Fourier transform (FFT) that can be applied to non-equidistantly sampled data in time/space or frequency domain. It is an approximative algorithm that allows to control the approximation error in such a way that machine precision is reached while keeping the algorithmic complexity in the same order as a regular FFT. The NFFT plays a major role in many signal processing applications and has been intensively studied from a theoretical and computational perspective. The fastest CPU implementations of the NFFT are implemented in the low-level programming languages C and C++ and require a compromise between code generalizability, code readability, and code efficiency. The programming language Julia promises new opportunities in optimizing these three conflicting goals. In this work we show that Julia indeed allows to develop an NFFT implementation, which is completely generic, dimension-agnostic and requires about 2--3 times less code than the other famous libraries NFFT3 and FINUFFT while still being one of the fastest NFFT implementations developed to date.
\end{abstract}




\definecolor{bg}{rgb}{0.95,0.95,0.95}

\algblock{ParFor}{EndParFor}
\algnewcommand\algorithmicparfor{\textbf{parfor}}
\algnewcommand\algorithmicpardo{\textbf{do}}
\algnewcommand\algorithmicendparfor{\textbf{end\ parfor}}
\algrenewtext{ParFor}[1]{\algorithmicparfor\ #1\ \algorithmicpardo}
\algrenewtext{EndParFor}{\algorithmicendparfor}

\algblock{Lock}{EndLock}
\algnewcommand\algorithmiclock{\textbf{lock}}
\algnewcommand\algorithmicendlock{\textbf{end\ lock}}
\algrenewtext{Lock}{\algorithmiclock}
\algrenewtext{EndLock}{\algorithmicendlock}

\section{Introduction}

The Fourier transform plays an important role in many mathematical applications in particular those involving signal processing. Convolutions in time/spatial domain can be expressed as multiplications in frequency domain, which is often faster and allows one to invert convolutions in analytical form. In practice, signals are usually discrete and of finite length such that the discrete Fourier transform (DFT) needs to be applied. It maps a number of $N$ samples in time/spatial domain to $N$ samples in frequency domain and requires ${\cal O}(N^2)$ operations, which is often too expensive to apply the DFT in its ordinary form. The fast Fourier transform (FFT)  \cite{cooley1965algorithm} allows one to carry out the DFT in only ${\cal O}(N\log N)$ operations and enables many applications for which the DFT would be too expensive. Its high impact in various applications makes the FFT one of the most important numerical algorithms developed in the 20th century. One of the fastest FFT software libraries is the FFTW \cite{FFTW05}, which is used in most scientific applications.

An important limitation of the FFT is that it requires the input and ouput signals to be equidistantly sampled. This makes it unusable for some important applications, e.g., in optical coherence tomography \cite{hillmann2009using} and magnetic resonance imaging \cite{fessler2007nufft,knopp2007note}. In this work we focus on the case, where the signal in one domain is sampled in a non-equidistant manner. In such cases, simple interpolation, to map from non-equidistant to equidistant points, leads to large numerical errors, which intuitively can be explained by the fact that local (interpolation) errors lead to global errors in the reciprocal domain. This motivated the development of the non-equidistant fast Fourier transform (NFFT), which allows to keep the approximation error below the floating point precision and thus can be used like an exact algorithm in most circumstances. It was developed in the 1990s and 2000s by various researchers \cite{dutt1993fast,beylkin1995fast,anderson1996rapid, steidl1998note,ware1998fast,potts2001fast} with a special focus on developing error bounds that allow to predict the approximation error in dependence of NFFT hyperparameters. One important ingredient of the NFFT is the use of a window function, which is convolved with the irregular data and sampled at equidistant points. In the past, several window functions have been proposed of which the Kaiser-Bessel function provides the highest accuracy, since it is well localized in time/spatial and frequency domain. Quite recently, new error bounds have been derived for different window functions \cite{potts2021uniform} including the Kaiser-Bessel window \cite{potts2021continuous}.

Beside this theoretical work, there is also a strong demand for high-quality fast software packages implementing the NFFT. While the algorithm itself has a  rather simple  mathematical notation, a straight forward textbook implementation would suffer from suboptimal performance and  be orders of magnitude slower than a tuned software library. For this reason, several high-performance NFFT libraries like NFFT3 \cite{keiner2009using} and FINUFFT \cite{barnett2019parallel} have been developed in the past. NFFT3 was developed in the late 2000s, is written C and supports multi-threading. In addition to the NFFT it offers algorithms for the non-equidistant fast cosine transform (NDCT) and the non-equidistant fast sine transform (NDST). Two important features of NFFT3 are the ability to apply the algorithm to arbitrary dimensional signals and the ability to cache window function evaluations for faster computation. FINUFFT was developed in the late 2010s, is written in C++ and provides two major innovations: First, it uses a new window function that can be evaluated in a  fast manner. Second, it uses a new multi-threading approach for the adjoint NFFT using a block-partitioning strategy. In combination with highly tuned C++ code, these two innovations make FINUFFT one of the fastest NFFT implementations developed to date. The NFFT has also been implemented on graphical processing units (GPU) \cite{sorensen2008accelerating,kunis2012nonequispaced,knoll2014gpunufft,yang2018new,shih2021cufinufft}, which can give additional speedups over CPU implementations but is not in the scope of the current paper.


When developing a new NFFT software library, there is a certain design space to be explored. An implementation can
\begin{enumerate}
\item be generic, i.e., allowing for different floating point types (\texttt{Float32}/\texttt{Float64}),
\item be dimension-agnostic,
\item allow changing the window function,
\item allow changing the precomputation strategy,
\item be fast/multi-threaded,
\item be readable/maintainable,
\item be re-usable/binding friendly.
\end{enumerate}
Since not all of these goals can be achieved at the same time, certain compromises must be made, many of which also depend on the programming language being used. Both NFFT3 and FINUFFT are implemented in C/C++ and made the following design decisions:
\begin{itemize}
    \item Both libraries are partly generic, i.e., they allow for \texttt{Float32} and \texttt{Float64} floating point types. To avoid manual code duplication, the number type is defined using a C macro allowing for automatic code copies and textual replacements. In fact, most of the NFFT3 code is written in macro form to avoid code duplication. 
    \item It is difficult to achieve a fast and dimension-independent implementation in C at the same time. This is because a static dimension, known at compile time, allows the compiler to generate much more efficient machine code. In particular nested for loops slow down code execution significantly if they are emulated in a dimension-agnostic fashion during runtime. Both libraries solve this problem by independent and redundant implementation of specific dimensions (NFFT3: 1D--5D, FINUFFT: 1D--3D). NFFT3 has an additional dynamic and slow fallback for higher dimensions while FINUFFT is restricted to 1D--3D. Thus, speed and dimension-agnosticity can be combined in C but this decreases the code readability/maintainability.
    \item Hardcoding the window function can lead to additional speedups since it avoids a dynamic dispatch in a hot code path. For this reason the window function is a compile time option in NFFT3 while it is hardcoded in FINUFFT.
\end{itemize}
The purpose of this work is to develop a new NFFT package that makes less trade-offs than its C/C++ counterparts. Our package, named NFFT.jl, is implemented in the programming language Julia \cite{bezanson2017julia}, which was first published in 2012 and has since then been developed into a popular programming language for scientific computing. Julia code can be both high-level like Matlab/Python and low-level like C/C++. The resulting machine code is usually as fast as comparable C/C++ machine code. This is achieved by a proper language design and the use of a just-in-time (JIT) compiler that allows to dynamically generate efficient machine code during runtime. Nowadays, Julia also supports multi-threading, which originates back to the prototype published in \cite{knopp2014experimental}.


This paper outlines the key implementation aspects of NFFT.jl and makes an in-depth performance analysis compared to NFFT3 and FINUFFT. Since the architecture of NFFT.jl is rather flexible, we were able to implement several features that are only available in either NFFT3 or FINUFFT. This allows us to compare and benchmark certain implementation strategies within the same code base. We also make algorithmical improvements to the strategies proposed in \cite{barnett2019parallel}, making NFFT.jl one of the fastest NFFT implementation developed to date.



%





\section{Mathematical Description}

We start with a mathematical description of the NFFT. Throughout this work, we stick to a multi-dimensional formulation and keep the notation close to the one used in \cite{keiner2009using}. Before discussing the NFFT, we first introduce the underlying mathematical transform to be computed: the non-equidistant discrete Fourier transform (NDFT).

\subsection{NDFT}

We let $\bm{N} \in \mathbb{N}^D$ be the size of the NDFT in the equidistant sampling domain and $D \in \mathbb{N}$ be its dimensionality. 
We further use the multi-dimensional index set
\begin{equation}
 I_{\bm{N}} := \mathbb{Z}^D \cap \prod_{d=1}^D \left[-\frac{N_d}{2},\frac{N_d}{2}\right) 
\end{equation}
representing all equidistant sampling points. The subindex $n_d$ covers $-\frac{N_d}{2}, \dots, \frac{N_d}{2}-1$ for even $N_d$ and $-\frac{N_d-1}{2}, \dots, \frac{N_d-1}{2}$ for odd $N_d$. The corresponding signal is denoted by $f_{\bm{n}} \in \mathbb{C}$, $\bm{n} \in I_{\bm{N}}$. 
In the frequency domain, the signal is sampled at non-equidistant sampling points $\bm{k}_j = \left( k_{d,j} \right)_{d=1}^{D} \in \mathbb{T}^D, j=1,\dots, J$ with $J \in \mathbb{N}$ and $\mathbb{T} := [-1/2,1/2)$. The resulting Fourier coefficients are denoted by $\hat{f}_{j} \in \mathbb{C}$, $j=1,\dots,J$. The (direct) NDFT is then defined as
\begin{align} \label{eq:NDFT}
   \hat{f}_j := \sum_{ \bm{n} \in I_{\bm{N}}} f_{\bm{n}} \, \mathrm{e}^{-2\pi\mathrm{i}\,\bm{n}\cdot\bm{k}_j}, \quad j=1,\dots, J & & \textit{(equidistant to non-equidistant)}, 
\end{align}
where $\bm{n}\cdot\bm{k}_j$ is the standard inner product between the two sampling points. This transformation is also known as the type-2 NDFT \cite{barnett2019parallel}. 

\begin{remark}
There are different conflicting definitions of the NDFT that consider the input signal to be sampled either in frequency \cite{keiner2009using} or in spatial domain \cite{fessler2003nonuniform}. Although we largely follow the notation of \cite{keiner2009using} we decided to switch frequency and spatial domain to make the NDFT consistent with the DFT, which is commonly defined to map from spatial to frequency domain. We note that this change only affects notation and that the computations remain the same.
\end{remark}

The NDFT has an associated adjoint that can be expressed as
\begin{align} \label{eq:NDFTH}
	y_{\bm{n}} := \sum_{j = 1}^{J} \hat{f}_j \, \mathrm{e}^{2 \pi \mathrm{i} \, \bm{n} \cdot \bm{k}_j}, \quad \bm{n} \in I_{\bm{N}} & & \textit{(non-equidistant to equidistant)}.
\end{align}
This is also named the type-1 NDFT \cite{barnett2019parallel} and it is related to the backward DFT. 
On purpose, we used the new variable $y_{\bm{n}}$ instead of $f_{\bm{n}}$ since the adjoint NDFT is in general not the inverse of the NDFT. This only holds true for special cases, i.e.,  for equidistant sampling nodes and when introducing a suitable normalization factor.

The algorithmic complexity for a direct computation of the NDFT and its adjoint is ${\cal O}(J|I_{\bm{N}}|)$ where $|I_{\bm{N}}| := \prod_{d=1}^D N_d$. This is because it requires two nested for loops, one over the $J$ sampling points $\bm{k}_j$ and one over the $|I_{\bm{N}}|$ equidistant sampling points $\bm{n}$.

\subsection{NFFT}
We next introduce the fast realization of the NDFT.
There are different approaches for accelerating the NDFT all having in common to use some sort of approximation to enable the usage of one or several ordinary FFTs. In this work, we consider the most popular and widely used approach, which uses convolutions to grid the non-equidistant signal to an equidistant signal enabling FFT usage. The basic idea of this approach is to exploit the convolution theorem, which states that a convolution in frequency domain corresponds to a multiplication in spatial domain. To exploit this, one introduces an artificial convolution in the non-equidistant sampling domain with a window function $\hat{\varphi}$. This is corrected in the equidistant sampling domain by dividing by the window's inverse Fourier transform $\varphi$, i.e., applying a deconvolution. The key point is that the artificial convolution allows one to switch from equidistant to non-equidistant sampling points and makes it possible to first use the FFT to switch domains and then apply the resampling. Since the operation being applied  is not shift-invariant in the discrete and non-equidistant setting we name it \textit{resampling} to avoid confusing it with a regular continuous or discrete convolution. The grid on which the FFT is applied is chosen to be larger than the grid on which the equidistant input signal is sampled. To this end, a so-called oversampling factor $\sigma > 1$ is introduced and an oversampled grid of size $\bm{\widetilde{N}} = \sigma \bm{N}$ is considered. $\sigma$ impacts the accuracy of the NFFT and in practice it is chosen in the range $[1.25, 2]$ with $2$ being the common default value. Smaller values decrease the accuracy of the NFFT but reduce the memory requirement and are therefore commonly used for large transforms of high dimensionality.


In summary, the NFFT consists of three steps:
\begin{enumerate}
    \item resampling correction in equidistant domain
    \item fast Fourier transform
    \item resampling to map from equidistant to non-equidistant domain
\end{enumerate}

\subsubsection{Direct NFFT}

We next provide a short derivation of the NFFT and start with the
 $D$-dimensional window function $\hat{\varphi}: \mathbb{R}^D \rightarrow \mathbb{R}$. The latter is based on a one-dimensional version $\hat{\varphi}_\text{Base}: \mathbb{R} \rightarrow \mathbb{R}$ using the tensor product
\begin{equation}
    \hat{\varphi}(\bm{k}) = \hat{\varphi}(k_1, \dots, k_D) = \prod_{d=1}^D \hat{\varphi}_\text{Base}\left( \frac{\widetilde{N}_d}{m} k_d \right).
\end{equation}
Here, $m \ll N_d$ is the second hyperparameter of the NFFT that controls the width of the window function. The window function $\hat{\varphi}$ is chosen in such a way that it is even and well localized in spatial and frequency domain. 
Due to the tensor product structure of $\hat{\varphi}$, its inverse Fourier transform $\varphi$ can be expressed as a tensor product as well:
\begin{equation}
    \varphi(\bm{n}) = \varphi(n_1, \dots, n_D) = \prod_{d=1}^D  \frac{m}{\widetilde{N}_d} \varphi_\text{Base}\left( \frac{m}{\widetilde{N}_d} n_d \right) .
\end{equation}
Moreover,  $\varphi$ is real because of the even symmetry of $\hat{\varphi}$. Finally, we require $\varphi(\bm{n}) \neq 0$ for $\bm{n} \in I_{\bm{N}}$, which is fulfilled for the commonly used window functions.

The ansatz for deriving the NFFT is based on the inverse Fourier transform
\begin{equation} \label{eq:FTWindow}
\varphi(\bm{n})  = \int_{\mathbb{R}^D} \hat{\varphi}(\bm{k}) \mathrm{e}^{2\pi\mathrm{i}\,\bm{n}\cdot\bm{k}} \textrm{d}\,\bm{k} = \sum_{\bm{r} \in \mathbb{Z}^D} \int_{\mathbb{T}^D} \hat{\varphi}(\bm{k}+\bm{r}) \mathrm{e}^{2\pi\mathrm{i}\,\bm{n}\cdot(\bm{k}+\bm{r})} \textrm{d}\,\bm{k}.
\end{equation}
When considering $\bm{n}\in \mathbb{Z}^D$ we can exploit the periodicity of the complex exponential  ($\mathrm{e}^{2\pi\mathrm{i}\,\bm{n}\cdot(\bm{k}+\bm{r})} = \mathrm{e}^{2\pi\mathrm{i}\,\bm{n}\cdot\bm{k}}$ for $\bm{r}\in\mathbb{Z}^D$) and introduce the one-periodization of the window function $\tilde{\varphi}(\bm{k}) :=  \sum_{\bm{r} \in \mathbb{Z}^D} \hat{\varphi} (\bm{k}+\bm{r})$ yielding
\begin{equation*} 
\varphi(\bm{n})  =  \int_{\mathbb{T}^D} \tilde{\varphi}(\bm{k}) \mathrm{e}^{2\pi\mathrm{i}\,\bm{n}\cdot\bm{k}} \textrm{d}\,\bm{k}.
\end{equation*}
Applying the substitution $\bm{k} \rightarrow \bm{k} - \bm{k}'$ results in
\begin{equation}
    \varphi(\bm{n}) = \int_{\mathbb{T}^D} \tilde{\varphi}(\bm{k}-\bm{k}') \mathrm{e}^{2\pi\mathrm{i}\,\bm{n}\cdot(\bm{k}-\bm{k}')} \textrm{d}\,\bm{k}'. 
\end{equation}
Dividing by $\varphi(\bm{n})$ and $\mathrm{e}^{2\pi\mathrm{i}\,\bm{n}\cdot \bm{k}}$ yields
\begin{equation}
    \mathrm{e}^{-2\pi\mathrm{i}\,\bm{n}\cdot \bm{k}} = \frac{1}{\varphi(\bm{n})} \int_{\mathbb{T}^D} \tilde{\varphi}(\bm{k}-\bm{k}') \mathrm{e}^{-2\pi\mathrm{i}\,\bm{n}\cdot \bm{k}'} \textrm{d}\,\bm{k}'. 
\end{equation}
Now we approximate the integral on the right hand side using a rectangular quadrature rule with $\bm{\widetilde{N}}=\sigma \bm{N}$ sampling nodes yielding
\begin{equation} \label{eq:expApprox}
    \mathrm{e}^{-2\pi\mathrm{i}\,\bm{n}\cdot \bm{k}} \approx \frac{1}{|I_{\bm{\widetilde{N}}}|\varphi(\bm{n})} \sum_{\bm{l} \in I_{\bm{\widetilde{N}}}} \tilde{\varphi}(\bm{k}-\bm{l}\,\oslash\,\bm{\widetilde{N}}) \mathrm{e}^{-2\pi\mathrm{i}\,\bm{n}\cdot (\bm{l}\,\oslash\,\bm{\widetilde{N}})}
\end{equation}
where $\oslash$ denotes the element-wise division. This means, we can approximate a complex exponential sampled at any $\bm{k} \in \mathbb{T}^D$ using a sum of shifted complex exponentials. 

Inserting the approximation \eqref{eq:expApprox} into the NDFT \eqref{eq:NDFT} yields
\begin{equation} \label{eq:approxNFFT1}
   \hat{f}_j \approx \underbrace{\sum_{\bm{l} \in I_{\bm{\widetilde{N}}}} \tilde{\varphi}(\bm{k}_j-\bm{l}\,\oslash\,\bm{\widetilde{N}}) \underbrace{ \sum_{ \bm{n} \in I_{\bm{N}}} \underbrace{ \frac{f_{\bm{n}}}{|I_{\bm{\widetilde{N}}}|\varphi(\bm{n})} }_\text{resampling correction}   \mathrm{e}^{-2\pi\mathrm{i}\,\bm{n}\cdot (\bm{l}\,\oslash\,\bm{\widetilde{N}})} }_\text{DFT}}_\text{resampling}.
\end{equation}
One can see that the resampling (last step) allows us to use the FFT (second step) since the sampling nodes $\bm{l}\,\oslash\,\bm{\widetilde{N}}$ are now equidistant. To account for the resampling, the inner part first applies a resampling correction.

Until now, we have not yet made the algorithm any faster than a direct evaluation of the NDFT since the resampling requires ${\cal O}(J|I_{\bm{\widetilde{N}}}|)$ operations. To do so, we need to exploit that the window function $\hat{\varphi}$ is well localized and thus close to zero for most of the evaluations performed during the resampling. To exploit this formally, we truncate $\hat{\varphi}$ at $\pm \bm{m} \oslash \bm{\widetilde{N}}$ with $\bm{m} = \left( m \right)_{d=1}^{D}$ and define the truncated window function
\begin{equation} \label{eq:approxTrunc}
    \hat{\psi}(\bm{k}) := \begin{cases}
    \hat{\varphi}(\bm{k})  & \text{for}\; \bm{k} \in 
    \prod_{d=1}^{D} [-\frac{m}{\widetilde{N}_d},\frac{m}{\widetilde{N}_d}) \\
   0 & \text{otherwise} \\
     \end{cases}
\end{equation}
with support $\text{supp}\,\hat{\psi} = \prod_{d=1}^{D} [-\frac{m}{\widetilde{N}_d},\frac{m}{\widetilde{N}_d})$. In the same way $\hat{\psi}_\text{Base}$ is defined to be $\hat{\varphi}_\text{Base}$ truncated to $[-1,1)$. 
From \eqref{eq:approxNFFT1}, it can be seen that the resampling step is based on the one-periodization $\tilde{\varphi}$. To accelerate its computation, we thus require the corresponding one-periodization $\tilde{\psi}(\bm{k}) :=  \sum_{\bm{r} \in \mathbb{Z}^D} \hat{\psi} (\bm{k}+\bm{r})$. Replacing $\tilde{\varphi}(\bm{k})$ by $\tilde{\psi}(\bm{k})$ in \eqref{eq:approxNFFT1} yields
\begin{equation} \label{eq:approxNFFT2}
   \hat{f}_j \approx \sum_{\bm{l} \in I_{\bm{\widetilde{N}},m}(\bm{k}_j)} \tilde{\psi}(\bm{k}_j-\bm{l}\,\oslash\,\bm{\widetilde{N}}) \sum_{ \bm{n} \in I_{\bm{N}}}  \frac{f_{\bm{n}}}{|I_{\bm{\widetilde{N}}}|\varphi(\bm{n})}   \mathrm{e}^{-2\pi\mathrm{i}\,\bm{n}\cdot (\bm{l}\,\oslash\,\bm{\widetilde{N}})},
\end{equation}
with the multi-index set
\begin{equation}
I_{\bm{\widetilde{N}},m}(\bm{k}) :=  \left\lbrace \bm{l} \in I_{\bm{\widetilde{N}}} \,:\,  \exists p \in \mathbb{Z} \; (  - \bm{m} \oslash \bm{\widetilde{N}} \leq \bm{k} - \bm{l}\oslash \bm{\widetilde{N}} + p\bm{1} < \bm{m} \oslash \bm{\widetilde{N}} )  \right\rbrace, 
\end{equation}
where $\bm{1} = \left( 1 \right)_{d=1}^{D}$. This is the set of indices $\bm{l}$ for which $\bm{k}_j-\bm{l}\,\oslash\,\bm{\widetilde{N}} \in \text{supp}( \tilde{\psi} )$ and in turn other indices can be dropped without changing the result of the computation.  It has at most $2m^D$ elements, which is much less than the $\prod_{d=1}^D \widetilde{N}_d$ elements of $I_{\bm{\widetilde{N}}}$. In particular, the number of elements is independent of $\bm{\widetilde{N}}$.
For the implementation described in \cref{sec:implementation} we require an alternative formulation of the index set $I_{\bm{\widetilde{N}},m}(\bm{k})$ that separates the individual dimensions:
\begin{align*}
 I_{\bm{\widetilde{N}},m}(\bm{k}) & = \prod_{d=1}^{D} I_{\widetilde{N}_d,m}(k_{d}) \; \text{with}\; I_{\widetilde{N},m}(k) = \{ \omega_{n,m}(k,\ell) \,:\, \ell \in \left\lbrace 1, \dots, 2m\right\rbrace \}.
\end{align*}
Here, the function $\omega_{\widetilde{N},m}: \mathbb{T} \times \{1, \dots, 2m\} \rightarrow I_{\widetilde{N}}$ with
\begin{equation*}
   \omega_{\widetilde{N},m}(k,\ell) = (\lceil{} \widetilde{N} (k\;\text{mod}\; 1)  - m + \ell-1 \rceil \;\text{mod}\; \widetilde{N} ) - \frac{\widetilde{N}}{2}
\end{equation*}
allows us to calculate the index set directly taking account the index wrap due to the periodization of the window function $\tilde{\psi}$.

\begin{algorithm}
  \caption{Direct NFFT}
  \label{Alg:NFFT}
  \begin{tabular}{ll}
    Input: & $D, m\in \mathbb{N}$, $\bm{N}, \bm{\widetilde{N}}\in \mathbb{N}^D$, with $\bm{\widetilde{N}}=\sigma \bm{N}$ and $\sigma > 1$,\\
          & non-equidistant sampling points $\bm{k}_j \in \mathbb{T}^D, j=1,\dots, J$, \\
          & equidistantly sampled signal $f_{\bm{n}} \in \mathbb{C}, \bm{n} \in I_{\bm{N}}$
  \end{tabular}
  \vspace{5pt}
  \begin{algorithmic}[1] 
      \For{$\bm{n} \in I_{\bm{\widetilde{N}}}$}
      \State $g_{\bm{n}} = \begin{cases}
    \frac{f_{\bm{n}}}{|I_{\bm{\widetilde{N}}}|\varphi(\bm{n})} & \text{for}\; \bm{n} \in I_{\bm{N}\vphantom{\widetilde{N}}} \\
   0 & \text{for}\; \bm{n} \in I_{\bm{\widetilde{N}}} \backslash  I_{\bm{N}\vphantom{\widetilde{N}}} \\
     \end{cases}$ \Comment{\textit{resampling correction}}
     \EndFor
     \For{$\bm{l} \in I_{\bm{\widetilde{N}}}$}
    \State   $\hat{g}_{\bm{l}} = \displaystyle\sum_{\bm{n} \in I_{\bm{\widetilde{N}}}} g_{\bm{n}} \textrm{e}^{-2\pi \textrm{i} \bm{n} \cdot (\bm{l} \,\oslash\, \bm{\widetilde{N}})}$  \Comment{\textit{FFT}}
    \EndFor
    \For{$j=1,\dots, J$} 
    \State $\displaystyle \hat{f}_j = \sum_{\bm{l} \in I_{\bm{\widetilde{N}},m}(\bm{k}_j)} \hat{g}_{\bm{l}}\,  \tilde{\psi} (\bm{k}_j -  \bm{l} \,\oslash\, \bm{\widetilde{N}})$ \Comment{\textit{resampling}}
    \EndFor  
  \end{algorithmic}
  \begin{tabular}{ll} & \\
    Output: & non-equidistantly sampled signal $\hat{f}_j \in \mathbb{C}, j=1,\dots, J$ \\
    Complexity: & ${\cal O}( |I_{\bm{\widetilde{N}}}|\log |I_{\bm{\widetilde{N}}}| + m^D J)$
\end{tabular}
\end{algorithm}

The direct NFFT is summarized in \cref{Alg:NFFT}:
\begin{enumerate}
    \item In the first step, the resampling correction is applied. To this end, the input data $f_{\bm{n}}$ is divided by the inverse Fourier transform of the window function $\varphi(\bm{n})$ for each $\bm{n} \in I_{\bm N}$. 
    The output of the resampling correction is stored in a new temporary vector $\bm{g} = \left( g_{\bm{n}} \right)_{\bm{n} \in I_{\bm{\widetilde{N}}}}$ that is defined on the fine grid $I_{\bm{\widetilde{N}}}$. The first step of the NFFT thus also applies the zero padding necessary for the in-place FFT in the second step of the algorithm.
    \item In the second step, an ordinary $D$-variate FFT is applied. It operates in-place but to keep the mathematical notation sound, we introduce the output vector $\hat{\bm{g}} = \left( \hat{g}_{\bm{l}} \right)_{\bm{l} \in I_{\bm{\widetilde{N}}}}$.
    \item In the third step, the resampling is applied.
\end{enumerate}



\subsubsection{Adjoint NFFT}

The adjoint NFFT is based on the same idea as the direct NFFT. Inserting \eqref{eq:expApprox} into \eqref{eq:NDFTH} and replacing $\tilde{\varphi}$ by $\tilde{\psi}$ yields
\begin{equation} \label{eq:approxNFFTH}
	y_{\bm{n}} \approx \underbrace{ \frac{1}{|I_{\bm{\widetilde{N}}}|\varphi(\bm{n})} \underbrace{\sum_{\bm{l} \in I_{\bm{N}}} \underbrace{\left(\sum_{j \in I^{\intercal}_{\bm{\widetilde{N}},m}(\bm{l})} \hat{f}_j\,  \tilde{\psi} (\bm{k}_j -  \bm{l} \,\oslash\, \bm{\widetilde{N}}) \right)}_\text{resampling} \, \mathrm{e}^{2 \pi \mathrm{i} \, \bm{n} \cdot (\bm{l} \,\oslash\, \bm{\widetilde{N}})} }_\text{adjoint DFT} }_\text{resampling correction}.
\end{equation}
Here, the index set $I^\intercal_{\bm{\widetilde{N}},m}(\bm{l})$ is defined as
\begin{equation}
I^\intercal_{\bm{\widetilde{N}},m}(\bm{l}) := \left\lbrace j \in \left\lbrace1,\dots, J\right\rbrace \,:\, \exists p \in \mathbb{Z} \; (  - \bm{m} \oslash \bm{\widetilde{N}} \leq  \bm{k}_j - \bm{l} \oslash \bm{\widetilde{N}} + p\bm{1} < \bm{m} \oslash \bm{\widetilde{N}} )  \right\rbrace 
\end{equation}
and again allows one to perform the inner resampling over only a subset of the original indices. One can also see that the adjoint NFFT applies the three steps in reverse to the direct NFFT.
The adjoint NFFT is summarized in \cref{Alg:NFFTAdjoint}: 
\begin{enumerate}
    \item In the first step, the non-equidistantly sampled signal $\hat{f}_j$ is convolved with the one-periodic window function $\tilde{\psi}$ yielding the equidistantly sampled signal $\hat{g}_{\bm{l}}$. The sum is again restricted to the subset of indices $I^\intercal_{\bm{\widetilde{N}},m}(\bm{l})$ at which $\tilde{\psi}$ is non-zero. However, the adjoint resampling needs a different evaluation order than indicated by the summation. The reason is that it is very in-efficient to determine  $I^\intercal_{\bm{\widetilde{N}},m}(\bm{l})$ during the summation. Instead, one uses two for loops: the inner looping over $\bm{l} \in I_{\bm{\widetilde{N}},m}(\bm{k}_j)$ and the outer looping over the sampling points $\bm{k}_j$, $j=1,\dots, J$. This change of evaluation order means that the direct and the adjoint resampling have a very similar structure. The only difference is that the summation over $\bm{l}$ has no data dependency for the direct transform whereas the adjoint transform needs to perform additive vector updates on $\hat{\bm{g}}$, which cannot be performed concurrently for different nodes $\bm{k}_j$.
    \item The second step is a $D$-variate adjoint FFT.
    \item The third step is the resampling correction. Similar to the direct transform, only the subset of $\bm{g}$ on the grid $I_{\bm{N}}$ needs to be considered in this step.
\end{enumerate}

\begin{algorithm}[t] 
  \caption{Adjoint NFFT}
  \label{Alg:NFFTAdjoint}
  \begin{tabular}{ll}
    Input: & $D, m\in \mathbb{N}$, $\bm{N}, \bm{\widetilde{N}}\in \mathbb{N}^D$, with $\bm{\widetilde{N}}=\sigma \bm{N}$ and $\sigma > 1$,\\
          &  non-equidistant sampling points $\bm{k}_j \in \mathbb{T}^D, j=1,\dots, J$, \\
          & non-equidistantly sampled signal $\hat{f}_j \in \mathbb{C}, j=1,\dots, J$
  \end{tabular}
  \vspace{5pt}
  \begin{algorithmic}[1]
    \For{$\bm{l} \in I_{\bm{\widetilde{N}}}$} 
    \State $\displaystyle \hat{g}_{\bm{l}} = \sum_{j \in I^{\intercal}_{\bm{\widetilde{N}},m}(\bm{l})} \hat{f}_j\,  \tilde{\psi} (\bm{k}_j -  \bm{l} \,\oslash\, \bm{\widetilde{N}})$ \Comment{\textit{adjoint resampling}}
    \EndFor 

     \For{$\bm{n} \in I_{\bm{\widetilde{N}}}$}
    \State   $g_{\bm{n}} = \displaystyle\sum_{\bm{l} \in I_{\bm{\widetilde{N}}}} \hat{g}_{\bm{l}}  \textrm{e}^{2\pi \textrm{i} \bm{n} \cdot (\bm{l} \,\oslash\, \bm{\widetilde{N}})}$  \Comment{\textit{adjoint FFT}}
    \EndFor    
    
     \For{$\bm{n} \in I_{\bm N}$}
      \State $y_{\bm{n}} = \frac{g_{\bm{n}}}{|I_{\bm{\widetilde{N}}}|\varphi(\bm{n})}$ \Comment{\textit{adjoint resampling correction}}
     \EndFor
  \end{algorithmic}
  \begin{tabular}{ll} & \\
    Output: & equidistantly sampled signal $y_{\bm{n}} \in \mathbb{C}$ \\
    Complexity: & ${\cal O}(|I_{\bm{\widetilde{N}}}|\log |I_{\bm{\widetilde{N}}}| + m^D J)$
\end{tabular}
\end{algorithm}

\subsubsection{Matrix-Vector Notation}

The NFFT can be written in matrix-vector notation, which is helpful for conceptual understanding and can also be used for actual implementations (see discussion about GPU implementations in section \ref{Sec:Discussion}). In matrix-vector form, the NFFT can be expressed as 
\begin{equation*}
    \hat{\bm{f}} = \bm{A} \bm{f} = \bm{B} \bm{F} \bm{D} \bm{f}
\end{equation*}
where $\bm{A} = \left( \mathrm{e}^{-2\pi\mathrm{i}\,\bm{n}\cdot\bm{k}_j} \right)_{j=1,\dots,J;\bm{n} \in I_{\bm{N}}} \in \mathbb{C}^{J \times |I_{\bm{N}}|}$ is the NDFT matrix, $\bm{f} = \left( f_{\bm{n}} \right)_{\bm{n} \in I_{\bm{N}}} \in \mathbb{C}^{|I_{\bm{N}}|}$ is the input vector, and $\hat{\bm{f}} = \left( \hat{f}_j \right)_{j=1}^{J} \in \mathbb{C}^J$ is the output vector. The first NFFT step is the multiplication with the generalized diagonal matrix
\begin{equation*}
    \bm{D} =  \left(   \delta_{\bm{l},\bm{n}}  \frac{1}{|I_{\bm{\widetilde{N}}}| \varphi(\bm{n})}   \right)_{\bm{l} \in I_{\bm{\widetilde{N}}};\bm{n} \in I_{\bm{N}}} \in \mathbb{R}^{|I_{\bm{\widetilde{N}}}| \times |I_{\bm{N}}|}
\end{equation*}
where $\delta$ is the Kronecker symbol. The second step is the application of the Fourier matrix
\begin{equation*}
    \bm{F} = \left( \mathrm{e}^{-2\pi\mathrm{i}\,\bm{n}\cdot (\bm{l} \,\oslash\, \bm{\widetilde{N}})} \right)_{\bm{n} \in I_{\bm{\widetilde{N}}};\bm{l} \in I_{\bm{\widetilde{N}}}} \in \mathbb{C}^{|I_{\bm{\widetilde{N}}}| \times |I_{\bm{\widetilde{N}}}|}.
\end{equation*}
The last step is the multiplication with the sparse matrix
\begin{equation} \label{eq:Bmatrix}
    \bm{B} = \left(  \tilde{\psi} (\bm{k}_j -  \bm{l} \,\oslash\, \bm{\widetilde{N}}) \right)_{j=1,\dots,J;\bm{l} \in I_{\bm{\widetilde{N}}}} \in \mathbb{R}^{J \times |I_{\bm{\widetilde{N}}}|}
\end{equation}
having at most $(2m)^D J$ non-zero entries. The adjoint NFFT in matrix-vector form is obtained by reversing the order of the three matrices and applying the adjoint operator to each step individually:
\begin{equation*}
    \bm{y} = \bm{A}^\mathsf{H} \hat{\bm{f}}  = \bm{D}^\intercal \bm{F}^\mathsf{H} \bm{B}^\intercal \hat{\bm{f}} 
\end{equation*}

\subsubsection{Algorithmic Complexity}

The algorithmic complexity of the direct NFFT and its adjoint is 
\begin{align*}
  {\cal O}(\underbrace{|I_{\bm{\widetilde{N}}}|}_{\text{resampling correction}} + \underbrace{|I_{\bm{\widetilde{N}}}| \log |I_{\bm{\widetilde{N}}}|}_{\text{FFT}} + \underbrace{(2m)^D J}_{\text{resampling}}) &= {\cal O}(|I_{\bm{\widetilde{N}}}| \log |I_{\bm{\widetilde{N}}}| + m^D J) \\
   & = {\cal O}\!\left( |I_{\bm{N}}| \log |I_{\bm{N}}| + \log\!\left(\frac{1}{\varepsilon}\right) J \right). 
\end{align*}
Here, we assumed in the last step that $\sigma$ is constant (e.g., $\sigma = 2$), and that the accuracy $\varepsilon$ improves exponentially with the kernel parameter $m$. The latter is fulfilled for all commonly used window functions.

Beside this theoretical consideration, the actual performance of the NFFT and its individual steps highly depends  on the dimensionality and the size of the problem. The resampling correction step is the cheapest one and can usually be neglected in terms of computation time. In the most common setting ($J \approx |I_{\bm{N}}|$), the FFT is the second fastest operation and the resampling is the primary bottleneck being up to one order of magnitude slower. This is the reason why NFFT implementations usually put most efforts into optimizing the resampling as much as possible. However, in sparse sampling settings ($J \ll |I_{\bm{N}}|$) the FFT can become a dominant factor if the resampling is done properly. In multi-threading applications it is thus important to parallelize all parts of the NFFT. Otherwise even the resampling correction can become a bottleneck.

\section{Implementation} \label{sec:implementation}

After formulating the NFFT and its adjoint in mathematical notation we next focus on our software package NFFT.jl. Looking back to the design goals sketched in the introduction we put the highest priority on a generic, dimension-agnostic and fast/multi-threaded implementation. The second most important property is to keep the code readable and maintainable. Finally, we also want our implementation to be flexible and allow to change window functions and precomputation strategies with low effort. Reusability in different programming languages was not a top priority during the design of NFFT.jl but we sketch potential strategies for using NFFT.jl in programming languages other than Julia in  \cref{Sec:Discussion}.



\subsection{Example Usage}

A typical example usage of NFFT.jl for a 2D transformation is outlined below:
\begin{noerr}
\begin{minted}[mathescape, linenos, 
bgcolor=bg, %frame=lines, 
numbersep=5pt, gobble=0, framesep=2mm]{julia}
using NFFT

D = 2                             # dimensionality
J = 32*32                         # number of sampling points
N = (32, 32)                      # input signal size
k = rand(D, J) .- 0.5             # sampling points in [-0.5,0.5]^D

p = NFFTPlan(k, N; m=4, σ=2)      # create the NFFT plan

f = randn(Complex{Float64}, N)    # signal to be transformed
fHat = p * f                      # compute direct NFFT
y = adjoint(p) * fHat             # compute adjoint NFFT
\end{minted}
\end{noerr}
For simplicity, the input signal and the sampling nodes are initialized with random numbers. Based on the signal size \texttt{N} (line 5) and the sampling nodes \texttt{k} (line 6), an \texttt{NFFTPlan} object is created in line 8. The constructor takes care of allocating all necessary memory for temporary arrays and performing precomputations to make later transformations as fast as possible. This precomputation approach is very common in scientific programming. The actual NFFT's are applied in line 11 (direct) and 12 (adjoint). Since the NFFT can be interpreted as a matrix-vector multiplication, NFFT.jl uses the method \texttt{*} to express the transformation. The method \texttt{adjoint} creates a lazy wrapper type that allows \texttt{*} to call the adjoint transformation. In this way the syntax is very close to the mathematical notation. The chosen interface also matches the common naming scheme and coding pattern used for linear transformations in the Julia ecosystem.

In our example, the output vector was allocated within the \texttt{*} method. To avoid this allocation one can use the interface
\begin{noerr}
\begin{minted}[mathescape, linenos,
bgcolor=bg, %frame=lines,
numbersep=5pt, gobble=0, framesep=2mm]{julia}
mul!(fHat, p, f)
mul!(y, adjoint(p), fHat)
\end{minted}
\end{noerr}
\noindent which allows one to pass the output vector as the first argument. Internally, \texttt{*} is implemented as a small wrapper around \texttt{mul!}. Again, \texttt{mul!} is a standard method in Julia to express in-place linear transformations. Finally, there is also a high-level interface
\begin{noerr}
\begin{minted}[mathescape, linenos, 
bgcolor=bg, %frame=lines,
numbersep=5pt, gobble=0, framesep=2mm]{julia}
nfft(k, f; m=4, σ=2) 
nfft_adjoint(k, N, fHat; m=4, σ=2) 
\end{minted}
\end{noerr}
\noindent that automatically creates a plan before calling the low-level NFFT functions. This is convenient if the NFFT is applied only once. 

\subsection{Memory Management}

\texttt{NFFTPlan} is a struct holding several temporary arrays required by the NFFT. In particular it holds:
\begin{itemize}
    \item the temporary vector $\hat{\bm{g}}$.
    \item the forward and backward plan of the inner FFT, i.e., data structures formed during precomputation to facilitate the computation of FFTs.
    \item different index and data vectors related to the resampling (see \cref{Sec:SubSecConvolution}) and the resampling correction (see \cref{Sec:SubSecDeconvolution}).
\end{itemize}
The concrete number of data that the plan holds depends on the precompu\-tat\-ion strategy (see \cref{Sec:Precomputation}). Furthermore, the block-partitioning strategy discussed in section \cref{Sec:Blocking} also needs additional memory.

\subsection{Generic Types}

One important design goal of NFFT.jl is the ability to use different number types. In Julia, this can be done by introducing a type parameter \texttt{T} that can take on any floating point number type. We restrict all real data types (like the input nodes \texttt{k} and the precomputed window function entries) to be of type \texttt{T} while complex values like the input and output vectors of the NFFT are of type \texttt{Complex\{T\}}. This forcing of a common number type is important for optimal performance, as the need for type promotion is removed. The type \texttt{T} is encoded in the \texttt{NFFTPlan} type and allows the JIT compiler to generate dedicated machine code for each number type. Index types are all stored as 64-bit integers, which is the native integer number type on most modern CPUs. In principle a flexibilization of the index type is straight forward but this will only be considered if the current restriction to 64-bit integers poses a real problem.

\subsection{Dimension-Agnostic Implementation}

The second important design goal of NFFT.jl is to be dimension-agnostic. This is achieved by an integer parameter \texttt{D}, which allows us to encode the dimension in the \texttt{NFFTPlan}. This ensures the generation of dedicated machine code for each dimension. For instance, the signal size $\bm{N}$ is stored as a fixed size tuple \texttt{N::NTuple\{D,Int64}\}.
Besides these storage aspects, the real challenge is to implement the actual calculations in a dimension-agnostic way. Two possible solutions to tackle this challenge are:
\begin{enumerate}
 \item Write dedicated implementations for each dimension $D$.
 \item Implement multi-dimensional loops using iterators that treat $D$ as a runtime parameter.
\end{enumerate}
The first solution ensures maximum performance but leads to code duplication. This is problematic since code changes need to be done in all copies of the code thus increasing the maintainability cost. The second solution can be implemented with less code and without any duplication but it has the downside of being much slower because the compiler cannot emit fast machine code. Because of these downsides we considered neither of these two solutions. Instead, we exploit that Julia's architecture enables the implementation of fast and generic dimension-agnostic Cartesian iterators. During the development of Julia, two different Cartesian iterator types have been established. The first option named \texttt{CartesianIndices} is iterator-based and allows one to iterate over a \texttt{D}-dimensional array \texttt{A} with
\begin{minted}[mathescape, linenos, 
bgcolor=bg, %frame=lines,
numbersep=5pt, gobble=0, framesep=2mm]{julia}
for i in CartesianIndices(A)
  A[i] = ...
end
\end{minted}
Since the dimension \texttt{D} is a parameter of the array \texttt{A}, the JIT compiler is able to generate efficient code for \texttt{CartesianIndices}-based for loops. The second option, which was developed much earlier, is a macro-based solution available in the \texttt{Base.Cartesian} module. A dimension-agnostic loop using \texttt{Base.Cartesian} is formulated as
\begin{minted}[mathescape, linenos, 
bgcolor=bg, %frame=lines,
numbersep=5pt, gobble=0, framesep=2mm]{julia}
@nloops $D i A begin
    @nref $D A i = ...
end
\end{minted}
and generates the following code for \texttt{D = 3}:
\begin{minted}[mathescape, linenos, 
bgcolor=bg, %frame=lines,
numbersep=5pt, gobble=0, framesep=2mm]{julia}
for i_3 = axes(A, 3)
    for i_2 = axes(A, 2)
        for i_1 = axes(A, 1)
            A[i_1, i_2, i_3] = ...
        end
    end
end
\end{minted}
This looks exactly like a hand-written implementation for a given dimensionality with the important difference, that the code is automatically specialized for the array dimension \texttt{D}.
In NFFT.jl, we mainly use the \texttt{CartesianIndices}-based solution, since it is easier to read and maintain. However, in cases where more flexibility and control is needed, we use the \texttt{Base.Cartesian} macros. In particular, we use it in situations where the inner and the outer for loop are handled differently. This allows us to apply multi-threading only to the outer for loop.


\subsection{Resampling Correction} \label{Sec:SubSecDeconvolution}
The resampling correction (and its adjoint) is a simple element-wise operation. Due to oversampling, the input and output array do not have the same size but the iteration only runs over $I_{\bm{N}}$. To this end, we use  \texttt{Base.Cartesian} and multi-thread the most outer for loop.

One important implementation detail of the resampling correction is to take the indices of the later FFT into account. The FFT is usually defined with a sum running from $0$ to $\widetilde{N}_d-1$ while the FFT within the NFFT needs to be  performed on the grid $I_{\bm{\widetilde{N}}}$ with indices running from $-\frac{\widetilde{N}_d}{2}$ to $\frac{\widetilde{N}_d}{2}-1$. Accordingly, an \texttt{fftshift} needs to be performed. To avoid having to perform an extra step, we integrate the index mapping into the resampling correction step and thus only need to touch each point once. 

The primary bottleneck of the resampling correction is the evaluation of the function $\varphi$. While this can often be neglected because the resampling is much more expensive, there are situations, such as 1D transforms with $J\ll |I_{\bm{N}}|$, where the resampling correction can become a bottleneck. Therefore we perform two optimizations:
\begin{enumerate}
    \item We cache $\varphi$ so that it can be reused when applying many NFFTs to different data. Here we can exploit the tensor product structure of $\varphi$ and cache the vectors
    \vspace{-0.2cm}\begin{equation*}
    \bm{\beta}^\text{tensor}_d = \left( \beta^\text{tensor}_{n_d,d} \right)_{n_d \in I_{N_d}}, \quad \beta^\text{tensor}_{n_d,d} =  \frac{1}{m\,\varphi_\text{Base}\left( \frac{m}{\widetilde{N}_d} n_d \right)}, \quad d=1,\dots,D.
    \end{equation*} \vspace{-0.3cm}
    \\
    During the resampling correction we can then calculate $\frac{1}{|I_{\bm{\widetilde{N}}}|\varphi(\bm{n})}$ on the fly using:
    \vspace{0.3cm}
    \begin{algorithmic}[1]
    \For{$n_{D} \in I_{N_D}$}
      \State $\gamma_D \,\leftarrow\, \beta^\text{tensor}_{n_D,D}$
      \For{$n_{D-1} \in I_{N_{D-1}}$}
        \State $\gamma_{D-1} \,\leftarrow\, \gamma_D \, \beta^\text{tensor}_{n_{D-1},D-1}$
        \State $\cdots$
        \For{$n_1 \in I_{N_1}$}
          \State $\gamma_{1} \,\leftarrow\, \gamma_2 \, \beta^\text{tensor}_{n_{1},1}$
           \State  $g_{\bm{n}} \leftarrow  
    f_{\bm{n}} \gamma_1$
        \EndFor
        \State $\cdots$
      \EndFor
    \EndFor
  \end{algorithmic}
  \vspace{0.3cm}
  Here, we used the assignment notation $\leftarrow$ to indicate that the temporary variables $\gamma_d$ are updated during the loop. This form of precomputation requires only $\mathcal{O}(\sum_{d=1}^{D} N_d)$ memory (and evaluations of $\varphi$), whereas a full precomputation has a requirement of $\mathcal{O}(\prod_{d=1}^{D} N_d)$. During the actual resampling correction, there is no real downside, since there is just one additional multiplication in the inner loop, which in practice is not necessarily slower because the bottleneck is the load and store operation acting on $g_{\bm{n}}$ and $f_{\bm{n}}$ in line 8. For completeness NFFT.jl also has an option for full precomputation, which is useful in the GPU prototype that we sketch in section \cref{Sec:Discussion}. 
    \item While the number of evaluations of $\varphi$ is greatly reduced in the multi-dimen\-sional case, we observed a measurable influence for 1D transformations. We therefore optimize the precomputation of $\frac{1}{|I_{\bm{\widetilde{N}}}|\varphi(\bm{n})}$ by approximating $\varphi$ with a Chebyshev polynomial, which is much faster to evaluate than our default window (Kaiser-Bessel).
\end{enumerate}

\subsection{Fast Fourier Transform}

The FFT is usually not part of the NFFT implementation itself. Instead  one uses an existing high-performance FFT library. We chose the FFTW.jl package, which is a wrapper around the popular FFTW library \cite{FFTW05}. An alternative is to use the binary compatible implementation from Intel's Math Kernel Library (MKL), which can be changed via a runtime switch. Right now, NFFT.jl is hardcoded to use FFTW/MKL, while in principle it is possible to make this step exchangable using the interface package AbstractFFTs.jl.

The FFTW package provides a planner interface to split off precomputat\-ions and preallocations. In NFFT.jl, the \texttt{FFTWPlan} is integrated into the \texttt{NFFTPlan}, thus separating planning from computation. Since FFTW requires a dedicated plan for forward and backward transformation, both plans are precomputed and stored in the \texttt{NFFTPlan}.
The number of FFTW threads is matched to the number of threads used for the resampling and its correction within the NFFT. FFTW.jl allows one to use the thread-pool of the Julia runtime, which enables nested parallelism, i.e., if multiple NFFT/FFT are called in parallel from different threads, there will be no over-commitment breaking down the performance.

\subsection{Resampling} \label{Sec:SubSecConvolution}
The most important and most challenging operation of any NFFT implementation is the resampling. While the mathematical formula looks  straight forward, there can be slowdowns of more than one order of magnitude between textbook implementations and optimized ones. We next outline the most important aspects that should be taken into account.

\subsubsection{Loop Optimization} \label{sec:loopoptim}

To discuss the details of the implementation, we recapitulate the resampling steps, which are given by
\begin{align} \label{eq:SumOverNode}
\text{direct resampling:} \quad & \hat{f}_j = \sum_{\bm{l} \in I_{\bm{\widetilde{N}},m}(\bm{k}_j)} \hat{g}_{\bm{l}}\,  \tilde{\psi} (\bm{k}_j -  \bm{l} \oslash \bm{\widetilde{N}}), \quad j=1,\dots,J \\
\label{eq:SumOverNodeAdjoint} \text{adjoint resampling:} \quad & \hat{g}_{\bm{l}} = \sum_{j \in I^\intercal_{\bm{\widetilde{N}},m}(\bm{l})} \hat{f}_j\,  \tilde{\psi} (\bm{k}_j -  \bm{l} \,\oslash\, \bm{\widetilde{N}}), \quad \bm{l} \in I_{\bm{\widetilde{N}}}
\end{align}
Since the direct and adjoint transform have a very similar structure, we can share most of the concepts and code discussed in this section.
The operations \eqref{eq:SumOverNode} and \eqref{eq:SumOverNodeAdjoint} can either be memory bound because of the access to $\hat{g}_{\bm{l}}$, $\hat{f}_j$ and $\bm{k}_{j}$ or computational bound because of the calculation of $\tilde{\psi} (\bm{k}_j -  \bm{l} \oslash \bm{\widetilde{N}})$. In \cref{Sec:Blocking} and \cref{Sec:BlockingImpl} we discuss how to prevent cache misses, which substantially accelerates memory access time. In \cref{Sec:Precomputation} we discuss different precomputation strategies for accelerating the computation of $\tilde{\psi} (\bm{k}_j -  \bm{l} \oslash \bm{\widetilde{N}})$.

When both aspects are carefully taken into account one can perform additional loop optimizations. While these are often applied automatically, appropriately written code is required for the compiler to do so. The first optimization is the usage of an (immutable) fixed-size array. This allows the compiler to allocate arrays on the stack and keep them in CPU registers. Furthermore, it enables loop unrolling and allows the compiler to use SIMD (single instruction multiple data) instructions for parallel processing on a single core. In Julia it is possible to create fixed-size arrays during runtime using tuple types. We therefore create $D$  fixed-size vectors 
\begin{align*}
\bm{\tilde{\psi}}_{d,j}^\text{local} & := \left( \tilde{\psi}_{\ell,d,j}^\text{local} \right)_{\ell=1}^{2m}, \quad d=1, \dots, D \\
\tilde{\psi}_{\ell,d,j}^\text{local} & :=  \tilde{\psi}_\text{Base} \left( \frac{1}{m}\left(\widetilde{N}_d k_{d,j} - \omega_{\widetilde{N}_d,m}(k_{d,j},\ell) \right) \right)  
\end{align*}
and calculate \eqref{eq:SumOverNode} by
\begin{equation} \label{eq:InnerTensorStructure}
\hat{f}_j = 
\sum_{\ell_D=1}^{2m} \tilde{\psi}_{\ell_D,D,j}^\text{local}  \sum_{\ell_{D-1}=1}^{2m} \tilde{\psi}_{\ell_{D-1},D-1,j}^\text{local}  \cdots \sum_{\ell_1=1}^{2m} \tilde{\psi}_{\ell_1,1,j}^\text{local}\, \hat{g}_{\bm{l}}, \quad \bm{l} = \left( \omega_{\widetilde{N}_d,m}(k_{d,j},\ell_d)  \right)_{d=1}^D.
\end{equation}
The nested product calculation in \cref{eq:InnerTensorStructure} exploits the tensor product structure of the window function and results in a very simple inner loop (over $\ell_1$) that can be fully optimized by the compiler.  

\begin{remark}
Local caching of the window function using the vectors $\bm{\tilde{\psi}}_{d,j}^\text{local}$ should not be confused with the precomputation of $\tilde{\psi}$ discussed in \cref{Sec:Precomputation}. This is because the local vector $\bm{\tilde{\psi}}_{d,j}^\text{local}$ is calculated right before the resampling. Hence it is a local computation trick that helps the compiler to generate very efficient machine code. On the other hand, this caching can make precomputation unnecessary for larger dimensions. This is because the calculation of all non-zero entries of the $\bm{B}$ matrix -- which might require an expensive window evaluation -- has an arithmetic complexity of ${\cal O}(mD)$ while the sum in \eqref{eq:InnerTensorStructure} requires ${\cal O}(m^D)$ operations. With increasing $D$, the caching operation becomes negligible even for expensive window functions.
\end{remark}

\subsubsection{Block-Partitioning Motivation} \label{Sec:Blocking}

The non-equidistant nature of the sampling points $\bm{k}_j$ implies that there is no natural ordering of the sampling points in multiple dimensions. This means that subsequent sampling points $\bm{k}_j$, $\bm{k}_{j+1}$  can have a large distance in $\mathbb{T}^D$. If this is the case, they interact with largely separated regions in the vector $\hat{\bm{g}}$. Computational-wise, this leads to cache misses and degrades performance. This is especially a problem when using multiple threads, where cache misses can be avoided if closely located points are processed on the same CPU core.

A second and much more severe issue comes up within the adjoint resampling step. The direct resampling accesses $\hat{\bm{g}}$ only for reading, which can be done concurrently without changing the result of the computation. The adjoint, however, performs in-place additions acting on $\hat{\bm{g}}$ and results in race conditions when carried out in parallel. Locking the access to $\hat{\bm{g}}$ using a mutex could solve this problem but basically results in a serial execution since the access to $\hat{\bm{g}}$ is a large portion of the computation time. In practice, it even degrades the performance, since the CPU pipeline cannot be optimally used because of the unpredictable locking of the access to $\hat{\bm{g}}$.


Both of the aforementioned issues have been tackled in NFFT3 \cite{volkmer2012openmp} and FINUFFT \cite{barnett2019parallel}. Ref. \cite{volkmer2012openmp} proposed node sorting and 1D block-partitioning  as independent concepts for the NFFT3 library. Sorting is done with respect to the index $\bm{l}$ that is obtained by flooring $\bm{\widetilde{N}} \,\odot\, \bm{k}_j-m \bm{1}$ to $I_{\bm{\widetilde{N}}}$. Block-partitioning is performed by splitting $\hat{\bm{g}}$ in its 1D in-memory representation into $T$ regions, where $T$ is the number of threads. Then, all points acting on a specific block are determined and used during the actual computation. This approach achieves close to linear speedup with the number of threads when the density of the sampling points is close to uniform. It, however, yields sub-optimal scaling for non-uniform sampling density since the load is not well-balanced. This issue was solved in \cite{barnett2019parallel} where a more general block-partitioning strategy was proposed. The first idea is to use multi-dimensional blocks to improve the data locality. The second idea is to use a fixed number of points per block yielding better load balancing. Decoupling the number of blocks from the number of threads (i.e., use more blocks than threads) ensures that a single long-lasting thread does not slow down the entire computation. We note that a block partitioning scheme to exploit data locality has already been proposed in \cite{johnson2009convolution}, although no focus has yet been placed on parallel implementation in that work.

In NFFT.jl we followed the approach proposed in  \cite{barnett2019parallel} with the main difference that our blocks are smaller and all have the same size. For reference, we also implement a regular resampling without block-partitioning, which can be used to investigate the performance gains of block-partitioning. To the best of our knowledge, both NFFT3 and FINUFFT use no block-partitioning in the direct resampling but only in the adjoint resampling. We apply the concept to both operations.


\subsubsection{Block-Partitioning Implementation} \label{Sec:BlockingImpl}

We next describe block-par\-ti\-tion\-ing formally. The basic idea  is to split the domain $\mathbb{T}^D$ into $\bm{P} \in 2\mathbb{N}^D$ equally sized blocks
\begin{equation}
    R_{\bm{p}} := \prod_{d=1}^{D} \left[\frac{p_d}{P_d}, \frac{p_d+1}{P_d}  \right), \quad \bm{p} \in I_{\bm{P}}
\end{equation}
such that $\bigcup_{\bm{p} \in I_{\bm{P}}} R_{\bm{p}} = \mathbb{T}^D$. Based on this we can collect the indices $j=1,\dots,J$ of the nodes $\bm{k}_j$ in a block $R_{\bm{p}}$ by
\begin{equation}
    \Gamma_{\bm{p}} := \left\{ j\in \left\{1,\dots, J\right\} \,:\, \bm{k}_j \in R_{\bm{p}}\right\}.  
\end{equation}
Instead of iterating over $j=1,\dots,J$ with a single for loop, we can now use two nested for loops:
\begin{enumerate}
    \item an outer for loop over the blocks $\bm{p} \in I_{\bm{P}}$.
    \item an inner for loop over the node indices $j \in \Gamma_{\bm{p}}$.
\end{enumerate}
We next partition the index set $I_{\bm{\widetilde{N}}}$ into $\bm{P}$ blocks of size $\bm{Q} = \left\lceil{} \bm{\widetilde{N}} \oslash \bm{P} \right\rceil$, where the ceiling is performed in an element-wise fashion. Then, we calculate those indices $\bm{l} \in I_{\bm{\widetilde{N}}}$ that could be necessary in the resampling for the nodes in the block $\bm{p}$. This is the index set
\begin{align*}
I^\text{block}_{\bm{p},m} & := \bigcup_{\bm{k} \in  R_{\bm{p}}} I_{\bm{\widetilde{N}},m}(\bm{k}), 
\end{align*}
which has $|\bm{Q}+2\bm{m}|$ entries. 

Having defined the index set $I^\text{block}_{\bm{p},m}$ we have everything in place to define the block-partitioned resampling. We start with the direct resampling, which is summarized in \cref{Alg:BlockedConvolution}. The outer loop of the algorithm runs over the blocks (line 1). For each block $\bm{p}$, a local cache $\hat{\bm{q}}_{\bm{p}} = \left( \hat{q}_{\bm{n},\bm{p}} \right)_{\bm{n} \in I^\text{block}_{\bm{p},m} } \in \mathbb{C}^{\bm{Q}+2\bm{m}}$ is created and the data from $\hat{\bm{g}}$ is copied to $\hat{\bm{q}}$ (lines 2--4). Then, the inner loop over all nodes within $\Gamma_{\bm{p}}$ is carried out. This loop initializes $\hat{f}_j$ with zero (line 6) and then performs the inner resampling over $I_{\bm{\widetilde{N}},m}(\bm{k}_j)$ where now the local cache $\hat{\bm{q}}$ is used instead of $\hat{\bm{g}}$ (lines 7--9). Multi-threading the direct block-partitioned resampling is straight forward. The outer for loop over the blocks is run in parallel (indicated by \textbf{parfor}) and does not need to take data dependencies into account.

\begin{algorithm}[h] 
  \caption{Block-Partitioned Resampling}
  \label{Alg:BlockedConvolution}
  \begin{algorithmic}[1] 
    \ParFor{$\bm{p} \in I_{\bm{P}}$} \Comment{for each block}
      \For{$\bm{l}  \in I^\text{block}_{\bm{p},m}$}
      \State $\hat{q}_{\bm{l},\bm{p}} \leftarrow  \hat{g}_{\bm{n}}$  \Comment{init block}
     \EndFor
     \For{$j \in \Gamma_{\bm{p}}$}  \Comment{for each node in block}
       \State $\hat{f}_j \leftarrow  0$
       \For{$\bm{l} \in I_{\bm{\widetilde{N}},m}(\bm{k}_j)$} \Comment{inner resampling}
         \State $\hat{f}_j \leftarrow \hat{f}_j + \hat{q}_{\bm{l},\bm{p}} \,  \tilde{\psi}( \bm{k}_j - \bm{l} \,\oslash\, \bm{\widetilde{N}})$
       \EndFor
     \EndFor
    \EndParFor
  \end{algorithmic}
\end{algorithm}

The block-partitioned implementation of the adjoint resampling is summarized in \cref{Alg:BlockedConvolutionAdj}. In the first step (lines 1--3), $\hat{\bm{g}}$ is initialized with zeros. Then, the main outer loop over the blocks $\bm{p}$ is initiated. It first initializes the local caches $\hat{\bm{q}}$ with zero (lines 5--7). Then, for each node index $j$ in $\Gamma_{\bm{p}}$ all summands acting on the local block $\hat{\bm{q}}$ are added at the appropriate location (lines 9--11). Afterwards, the local cache needs to be added to the global vector $\hat{\bm{g}}$ (lines 14--16). To multi-thread the block-partitioned adjoint resampling, first the initial loop initializing $\hat{\bm{g}}$ with zero is run in parallel. Then, the main outer for loop over the blocks is multi-threaded. As a matter of fact, the data dependency caused by the in-place addition now needs to be taken into account. The first loop in lines 8--12 is unproblematic, since $\hat{\bm{q}}$ is thread-local and the inner loop is run in series. However, the second loop (lines 14--16), needs to be locked with a mutex, since otherwise different threads would simultaneously update values in $\hat{\bm{g}}$. While this locking could potentially slowdown computation, we note that in practice, the first loop (lines 8--12) has a much higher workload and therefore observed only a small scaling penalty. This is in full agreement with the observations made in \cite{barnett2019parallel}. 

\begin{algorithm}[ht!] 
  \caption{Block-Partitioned Adjoint Resampling}
  \label{Alg:BlockedConvolutionAdj}
  \begin{algorithmic}[1]
     \ParFor{$\bm{l}  \in I_{\bm{\widetilde{N}}}$}  
      \State $\hat{g}_{\bm{l}} \leftarrow 0$ \Comment{init output array}
      \EndParFor
    \ParFor{$\bm{p} \in I_{\bm{P}}$}      \Comment{for each block}
      \For{$\bm{l}  \in I^\text{block}_{\bm{p},m}$}    
      \State $\hat{q}_{\bm{l},\bm{p}} \leftarrow 0$ \Comment{init block}
     \EndFor
     \For{$j \in \Gamma_{\bm{p}}$}   \Comment{for each node in block}
       \For{$\bm{l} \in I_{\bm{\widetilde{N}},m}(\bm{k}_j)$} \Comment{inner resampling} 
         \State $\hat{q}_{\bm{l},\bm{p}} \leftarrow \hat{q}_{\bm{l},\bm{p}} + \hat{f}_j \,  \tilde{\psi}( \bm{k}_j - \bm{l} \,\oslash\, \bm{\widetilde{N}})$
       \EndFor
     \EndFor
     \Lock \Comment{critical section}
      \For{$\bm{l}  \in I^\text{block}_{\bm{p},m}$} 
      \State $\hat{g}_{\bm{l}} \leftarrow \hat{g}_{\bm{l}} +  \hat{q}_{\bm{l},\bm{p}}$ \Comment{add to output array}
      \EndFor
      \EndLock
    \EndParFor
  \end{algorithmic}
\end{algorithm}

\begin{remark}
In practice, the sets $I^\text{block}_{\bm{p},m}$ and $I_{\bm{\widetilde{N}},m}(\bm{k}_j)$ can be calculated efficiently. Both represent structured blocks within $I_{\bm{\widetilde{N}}}$ that can be represented by just start/end indices in each dimension. In some cases, an index wrap needs to be taken into account. This wrapping can be implemented efficiently by precalculating the wrapping indices for each block. In the actual implementation, all indices are shifted to positive indices with an additional index offset that needs to be taken into account during the calculation. The loop over $I_{\bm{\widetilde{N}},m}(\bm{k}_j)$ can be implemented without wrapping since $\hat{\bm{q}}$ contains extra padding, i.e.,  wrapping only needs to be taken into account in the for loop in lines 2--4 of the direct resampling and lines 14--16 of the adjoint resampling.
\end{remark}

\subsubsection{Window precomputation} \label{Sec:Precomputation}

Next we discuss different strategies to avoid large computational cost of the window function evaluation. In the literature (e.g., \cite{kunis2008time} and \cite{barnett2019parallel}), we found the following options:
\begin{enumerate}
    \item \textbf{No precomputation}\\ One can directly evaluate the window during the application of the resampling. Depending on the window, this can lead to very slow runtimes if the window evaluation is expensive.  
    \item \textbf{Full precomputation}\\ Another common strategy is to compute  all window entries prior to the transformation. There are different variants for this, either one can keep the indexing logic the same and just use the cached window entry if needed. Alternatively one can go one step further and represent the entire resampling as a multiplication with a sparse matrix $\bm{B}$, see \eqref{eq:Bmatrix}. This requires storing the indices of the non-zero entries in $\bm{B}$. 
    \item \textbf{Tensor product based precomputation}\\
    The tensor product approach is similar to full computation but calculates the window function in each direction separately. This is basically the same as using the local caches $\bm{\tilde{\psi}}_{d,j}^\text{local}$ with the exception that the caches are stored globally for all nodes $\bm{k}_j$.
    Pulling all window entries into the inner sum, as full precomputation would do, is in many situations slower, since it increases the memory bandwidth, while the local tensor caching approach (with and without precomputation) is CPU bound.

    \item \textbf{Linear interpolation based precomputation}\\
    A common pattern to speed up the evaluation of the window function is linear interpolation. To this end, during precomputation a lookup table
    \begin{align}
        \bm{\tilde{\psi}}^\text{linear} &:= \left( \tilde{\psi}_s^\text{linear} \right)_{s=0}^{S-1} \quad \text{with} \quad
        \tilde{\psi}_s^\text{linear} := \tilde{\psi}_\text{Base} \left( \frac{s}{S-2} \right)  
    \end{align}
    is created where $S$ is the number sampling points. During resampling one then needs to perform the lookup
    \begin{align} \label{eq:linInterp}
        \tilde{\psi}_\text{Base} \left( \frac{1}{m}\left(\widetilde{N}_d k_{d,j} -  l_d \right) \right) & \approx \tilde{\psi}_{\tilde{s}}^\text{linear}  + \alpha\, (\tilde{\psi}_{\tilde{s}+1}^\text{linear}-\tilde{\psi}_{\tilde{s}}^\text{linear}) 
    \end{align} 
    where 
    \begin{align*}
        \kappa &=  \left\vert  \frac{S-2}{m}\left(\widetilde{N}_d k_{d,j} -  l_d \right) \right\vert, \quad
        \tilde{s} = \left \lfloor{}  \kappa \right\rfloor, \quad
        \alpha =  \kappa -  \left \lfloor{}  \kappa \right\rfloor.
    \end{align*} 
    The sampling point $s=S-1$ laying outside the support of the window function is required since for $|\widetilde{N}_d k_{d,j} -  l_d | = m$, we have $\tilde{s} = S-2$ and in turn the lookup table $\bm{\tilde{\psi}}^\text{linear}$ is accessed at index $S-1$ in \eqref{eq:linInterp}. When choosing $S-2$ to be a multiple of $m$ (i.e., $S = mY + 2$, $Y \in \mathbb{N}$), it is possible to perform the calculation of $\kappa, \tilde{s}, \alpha$ only once and increase $\tilde{s}$ with a constant offset  when iterating $l_d$ through the index set $I_{\widetilde{N}_d, m}$.
    The error of the linear interpolation depends on $S$, i.e., the larger $S$, the smaller is the error. NFFT3 provides experimentally determined values for $S$, which keep the window approximation error for specific $m$ below the entire NFFT error. For instance, for the Kaiser-Bessel window they can be written as
    \begin{equation*}
     S = m 2^{c_{\text{min}(m,8)}} + 2, \quad \bm{C} = (c_m)_{m=1}^{8} = \begin{pmatrix} 3 & 7 & 9 & 14 & 17 & 20 & 23 & 24 \end{pmatrix}.
    \end{equation*}
    Performance-wise it turns out that this method works well for small $m<5$ but  larger $m$ require very large lookup tables. For instance, $m=8$ and the Kaiser-Bessel window requires $134217730$ entries (1 GB in double precision). This large amount of memory makes the operation much slower for large $m$ since the lookup table does not fit into cache anymore and the resampling basically needs to run over the entire look-up table with a spacing of $\frac{S-2}{m}$ in $2m$ steps.  
     \item \textbf{Polynomial approximation based precomputation}\\
 An alternative window approximation was proposed in \cite{barnett2019parallel}. It uses piecewise polynomial approximation with a high polynomial degree. The idea is to split the window into $2m$ parts and perform  polynomial approximation independently on each interval $\bigl[-1+(\ell-1)/m,-1+\ell/m\bigr)$, $\ell=1,\dots,2m$ by setting up the Vandermonde matrix $\bm{V}$ and the corresponding sampling vector $\bm{b}$:
 \begin{align*}
     \bm{V} &= \left( \left(-\frac{1}{2}+\frac{\vartheta}{\Theta-1} \right)^{\!z} \right)_{\vartheta=0,\dots,\Theta-1;z=0,\dots,Z-1}, \\
     \bm{b}_\ell &= \left(  \tilde{\psi}_\text{Base} \left( -1 + \frac{1}{m} \left( \ell -1 +\frac{\vartheta}{\Theta-1}\right)   \right)   \right)_{\vartheta=0}^{\Theta-1}.
 \end{align*}
The polynomial coefficients $\bm{\mu}_\ell = \left( \mu_{z,\ell} \right)_{z=0}^{Z-1}$ are then calculated by solving the least-squares problem $\Vert \bm{V} \bm{\mu}_\ell  - \bm{b}_\ell  \Vert_2 \overset{\bm{\mu}_\ell}{\rightarrow} \text{min}$. The reason to consider the least-squares problem is that we apply an oversampling and choose more sampling points than the polynomial degree $\Theta = 2Z$ to achieve smaller errors. We thus are in the approximation setting and not the interpolation setting. 
During resampling the polynomial approximation uses the lookup
 \begin{align*}
    \tilde{\psi}_\text{Base} \left( \frac{1}{m}\left(\widetilde{N}_d k_{d,j} -  \omega_{\widetilde{N}_d,m}(k_{d,j},\ell) \right) \right) & \approx \sum_{z=0}^{Z-1} \mu_{z,\ell}\,\zeta^z, \quad \ell = 1, \dots, 2m
\end{align*} 
where $\zeta = \widetilde{N}_d k_{d,j} - \left\lfloor{} \widetilde{N}_d k_{d,j} \right\rfloor -\frac{1}{2}$. By using $2m$ intervals, the position $\zeta$ is the same for all intervals and therefore only the weights $\mu_{z,\ell}$ need to change when iterating through $\ell = 1, \dots, 2m$.


Our polynomial approximation implementation is slightly different then the one implemented in FINUFFT. In particular, we sample the window function equidistantly on the real line, whereas FINUFFT uses a sampling of a square in the complex space that is tailored to the specific window function used in FINUFFT. In our experiments a polynomial degree of $Z-1 = 2m$ is sufficient to keep the window function approximation error below the NFFT error, while \cite{barnett2019parallel} reported $2m+3$ to be necessary. This might be due to the different sampling when setting up the approximation problem or the different window function being used. In contrast to FINUFFT, which hardcodes the polynomial coefficients for a specific set of $m$/$\sigma$ in a C file generated by a Matlab script, NFFT.jl sets up the coefficients $\mu_{z,\ell}$ up on the fly during precomputation. 
\end{enumerate}
\cref{tab:MemReqPre} summarizes the memory requirement for the different precomputation options. The least memory is required for no precomputation followed by polynomial approximation, which effectively always requires less than or equal to 272 entries taking into account that machine precision is reached for $m=8$. Full and tensor product precomputation are more expensive since they have a per node cost. For $D>1$ tensor precomputation requires much less memory then full precomputation. Linear interpolation is difficult to classify, since it can require less memory than full/tensor precomputation for small $m$ and large $J$ but also require more memory for large $m$ and small $J$.

\renewcommand{\arraystretch}{1.1}
\begin{table}[]
    \caption{Memory requirement for different window function precomputation strategies.}
    \label{tab:MemReqPre}
    \centering
\begin{tabular}{l|l}
\textbf{Precomputation}    & \textbf{Memory [\# values]} \\ \hline
 No     & - \\
 Full     & $(2m)^DJ$ \\
 Tensor     & $2mDJ$ \\
 Linear     & $S = m Y + 2$\\
 Polynomial & $2mZ$ 
\end{tabular}
\end{table}
\renewcommand{\arraystretch}{1.0}

We implemented all precomputation strategies except for no precomputation in NFFT.jl. They can be changed by passing the \texttt{precompute} option to the \texttt{NFFTPlan} and setting it to one of the enum values \texttt{FULL}, \texttt{TENSOR}, \texttt{LINEAR}, or \texttt{POLYNOMIAL}. The reason for not implementing no precomputation is that an efficient implementation of this option with exchangeable window function requires making the window function a type parameter of the \texttt{NFFTPlan}. Since direct evaluation is known to be slower than polynomial interpolation even for cheap windows (see the discussion in \cite{barnett2019parallel}), we refrained from implementing no precomputation to avoid an increase in software complexity. Tensor product precomputation is combined with polynomial approximation in NFFT.jl, i.e., the precomputation is accelerated with polynomial approximation when using the \texttt{TENSOR} option.

Having a look at the two reference libraries, NFFT3 supports the first four precomputation strategies, while FINUFFT supports only polynomial approximation.

\subsubsection{Window Size Considerations} \label{sec:WindowSize}

We defined the truncated window function $\hat{\psi}$ to have a support of $[-\frac{m}{\widetilde{N}_d}, \frac{m}{\widetilde{N}_d})$ along dimension $d$. This means that the sum within the resampling has $(2m)^D$ non-zero summands. When considering the support $[-\frac{m}{\widetilde{N}_d}, \frac{m}{\widetilde{N}_d} ]$, as commonly done in the literature, one needs to consider the special case that a node $\bm{k}_j$ lies exactly on the grid specified by $\bm{l} \oslash \bm{\widetilde{N}}$ and in turn the sum has $(2m+1)^D$ summands. In principle one could switch between $(2m+1)^D$ and $(2m)^D$ by checking if $\bm{k}_j$ lies on the grid. As a matter of fact, this would actually harm the performance, since an \texttt{if} statement in a hot loop would need to dynamically switch between $(2m)^D$ and $(2m+1)^D$, which could lead to CPU pipeline flushes.

Our solution mitigates the problem with the potential downside of a slightly higher error. This, however, is very unlikely since $\bm{k}_j$ would need to match a grid point exactly in floating point precision. An alternative to our approach is implemented in NFFT3. It considers $(2m+2)^D$ points and uses $\hat{\varphi}$ instead of $\hat{\psi}$ during window evaluation. This has the advantage that no unnecessary multiplications with zero are carried out, which would happen if $\hat{\psi}$ was sampled outside of its support. In turn, this method can lower the approximation error of the NFFT. In \cref{Sec:Results}, we investigate how much the choice of $(2m+2)^D$ points improves the accuracy and whether it is better to instead use $(2m)^D$ points and increment $m$, which results in the same number of operations but a different window shape.







\section{Materials and Methods}

In the remainder of the manuscript an extensive evaluation of NFFT.jl is carried out. Both accuracy and performance are investigated and compared to NFFT3 and FINUFFT.

In all examples $J = | I_{\bm{N}} |$ random sampling points $\bm{k}_j \in \mathbb{T}^D$  are chosen. This is a typical yet challenging use case and in particular the random nodes can be considered to be the worst-case scenario an NFFT implementation needs to tackle. We use 1D--3D datasets with $N_\text{1D} =  2^{18} = 262144$, $\bm{N}_\text{2D} = (512, 512)^\intercal$, $\bm{N}_\text{3D} = (64, 64, 64)^\intercal$. The oversampling parameter is chosen to be $\sigma = 2$ while the kernel size parameter $m$ is chosen between $3$ and $10$. For NFFT.jl and NFFT3, the Kaiser-Bessel window (Kaiser-Bessel function for $\varphi$ and the corresponding Fourier transform for $\hat{\varphi}$) is chosen. FINUFFT uses the \textit{exponential of semicircle} kernel.

All computations were performed on a computer with 1024 GB of main memory and an AMD EPYC 7702 CPU. NFFT.jl was used in version 0.13. FINUFFT (v2.1.0) and NFFT3 (v3.5.2) were applied using the Julia wrapper packages FINUFFT.jl and NFFT3.jl. Julia was used in version 1.8.2. FFTW is used by all libraries and the same option (\texttt{FFTW\_MEASURE}) was used such that the FFT computation time is the same. This has been verified by inspecting the inner NFFT timings, which can be accessed for all libraries. Benchmarks are performed by repeating the same computation several times (with a 120~seconds threshold) and using the minimum time of all runs. This can be viewed as an estimate for the lower bound that is reached with hot CPU caches and no other workload affecting the computation. The accuracy of the NFFT and its adjoint are determined by calculating the relative error $\frac{\Vert \bm{f}_\text{NDFT} - \bm{f}_\text{NFFT} \Vert_\infty}{\Vert \bm{f}_\text{NDFT} \Vert_\infty}$ where $\bm{f}_\text{NDFT}$ is the result vector when applying the NDFT and $\bm{f}_\text{NFFT}$ is the result vector of the NFFT.

The code for the entire analysis can be accessed in the \texttt{benchmark/paper} directory of the NFFT.jl GitHub repository, which can be found at: 
\url{https://github.com/JuliaMath/NFFT.jl}.


\section{Results} \label{Sec:Results}

\subsection{Accuracy Analysis}

\begin{figure}[t!]
  \centering
  \label{fig:a}\includegraphics[width=\textwidth]{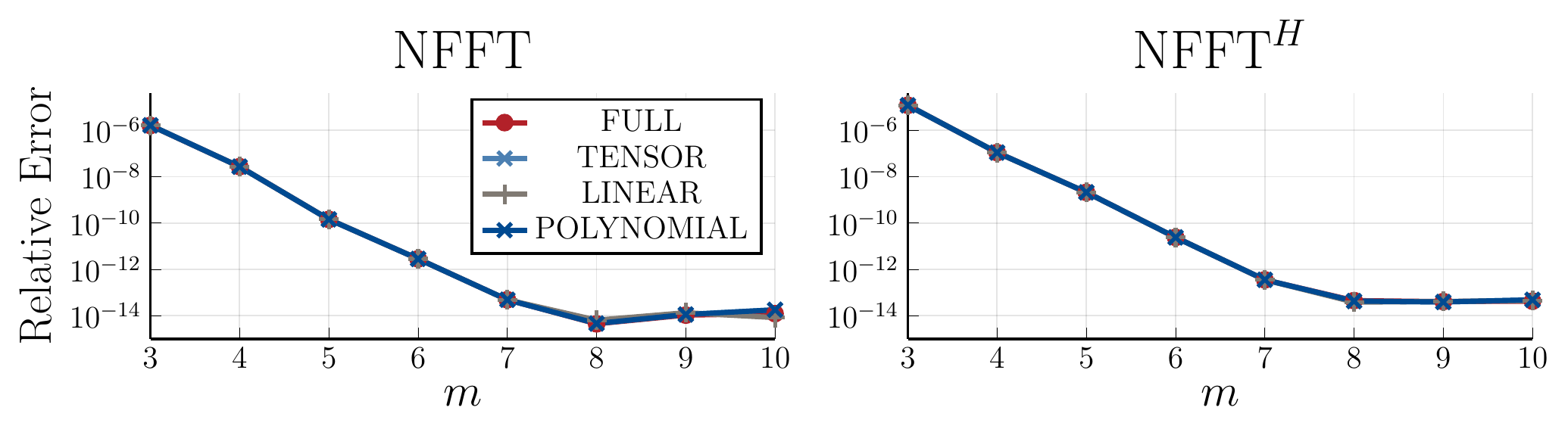}
  \vspace{-0.7cm}
  \caption{Relative error of NFFT.jl for different precomputation strategies, an oversampling factor of $\sigma = 2$ and window size parameters $m=3, \dots, 10$. Left shows the error of the direct transform while right shows the error of the adjoint transform. }
  \label{fig:accuracy1}
\end{figure}

We start the evaluation by looking at the accuracy of the NFFT.jl for the 2D dataset. The relative error of the direct and adjoint transform is shown for different precomputation strategies of the window function, an oversampling factor of $\sigma = 2$ and $m=3, \dots, 10$ in \cref{fig:accuracy1}. First of all one can see that the error decreases exponentially with the kernel parameter $m$. For the Kaiser-Bessel window the optimum accuracy is reached at $m=8$ at which point the error saturates. When comparing the different precomputation strategies one can observe almost no difference. This is remarkable since \texttt{FULL} does not apply a window approximation, while \texttt{TENSOR} (in NFFT.jl),  \texttt{LINEAR} and \texttt{POLYNOMIAL} do. This shows that by a proper selection of the window approximation parameters it is possible to keep this approximation error below the other approximation errors of the NFFT. 
Accuracy-wise, it thus does not matter which precomputation strategy is used.

\begin{figure}[t!]
  \centering
  \includegraphics[width=\textwidth]{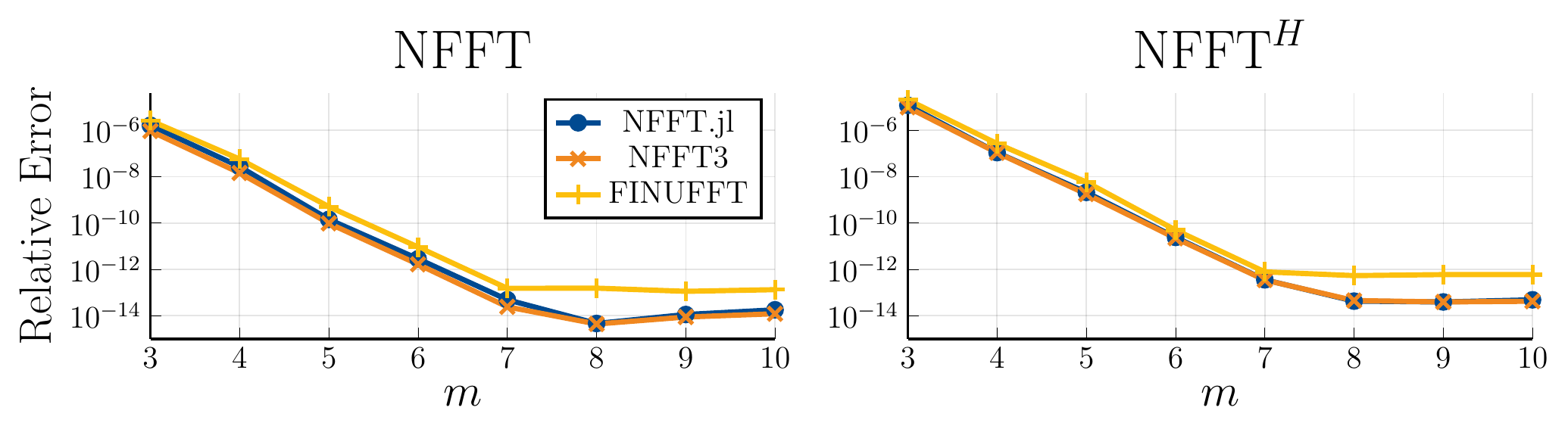}
  \vspace{-0.7cm}
  \caption{Relative error of NFFT.jl compared to NFFT3 and FINUFFT for an oversampling factor of $\sigma = 2$ and window size parameters $m=3, \dots, 10$. Left shows the error of the direct transform while right shows the error of the adjoint transform.}
  \label{fig:accuracy2}
\end{figure}

We next compare the accuracy between NFFT.jl, NFFT3 and FINUFFT. FINUFFT does not allow to directly set the parameters $\sigma$ and $m$ but instead uses a tolerance parameter $\varepsilon$ to derive $\sigma$ and $m$. By reading the source code and enabling debugging information, it was possible to select $\varepsilon$ such that, for a fixed value $\sigma=2$, the desired value $m$ was obtained. Due to an upper limit of $m=8$, larger $m$ were omitted for FINUFFT. 
The accuracy comparison is shown in \Cref{fig:accuracy2}. In a first qualitative comparison, one can see that the errors are very similar with NFFT3 being slightly more accurate than NFFT.jl for the direct NFFT, and NFFT.jl being more accurate than FINUFFT for both transformations. When looking at the quantitative values, the mean relative error ratio between NFFT.jl and NFFT3 is 1.767 for the direct transform and 1.106 for the adjoint transform. The mean relative error ratio between FINUFFT and NFFT.jl is 2.722 for the direct transform and 2.242 for the adjoint transform. In all cases, only $m=3, \dots, 7$ have been used for the calculation of the mean ratio. The differences in accuracy can be explained as follows. FINUFFT uses a different window function that is not as accurate as the Kaiser-Bessel window used in the other two implementations. The difference between NFFT.jl and NFFT3 is caused by NFFT.jl using a window of size $2m$ while NFFT3 uses a window of size $2m+2$ leading to a slightly lower window truncation error. However, the accuracy gain by increasing $m$ by one -- resulting in a window size of $2(m+1) = 2m + 2$ -- is much greater than the gain when using $2m+2$ points without changing the shape of the window. This justifies the choice of $2m$ points taken by NFFT.jl and FINUFFT. 


\subsection{Block-Size Investigation}

\begin{figure}[t!]
  \centering
  \includegraphics[width=\textwidth]{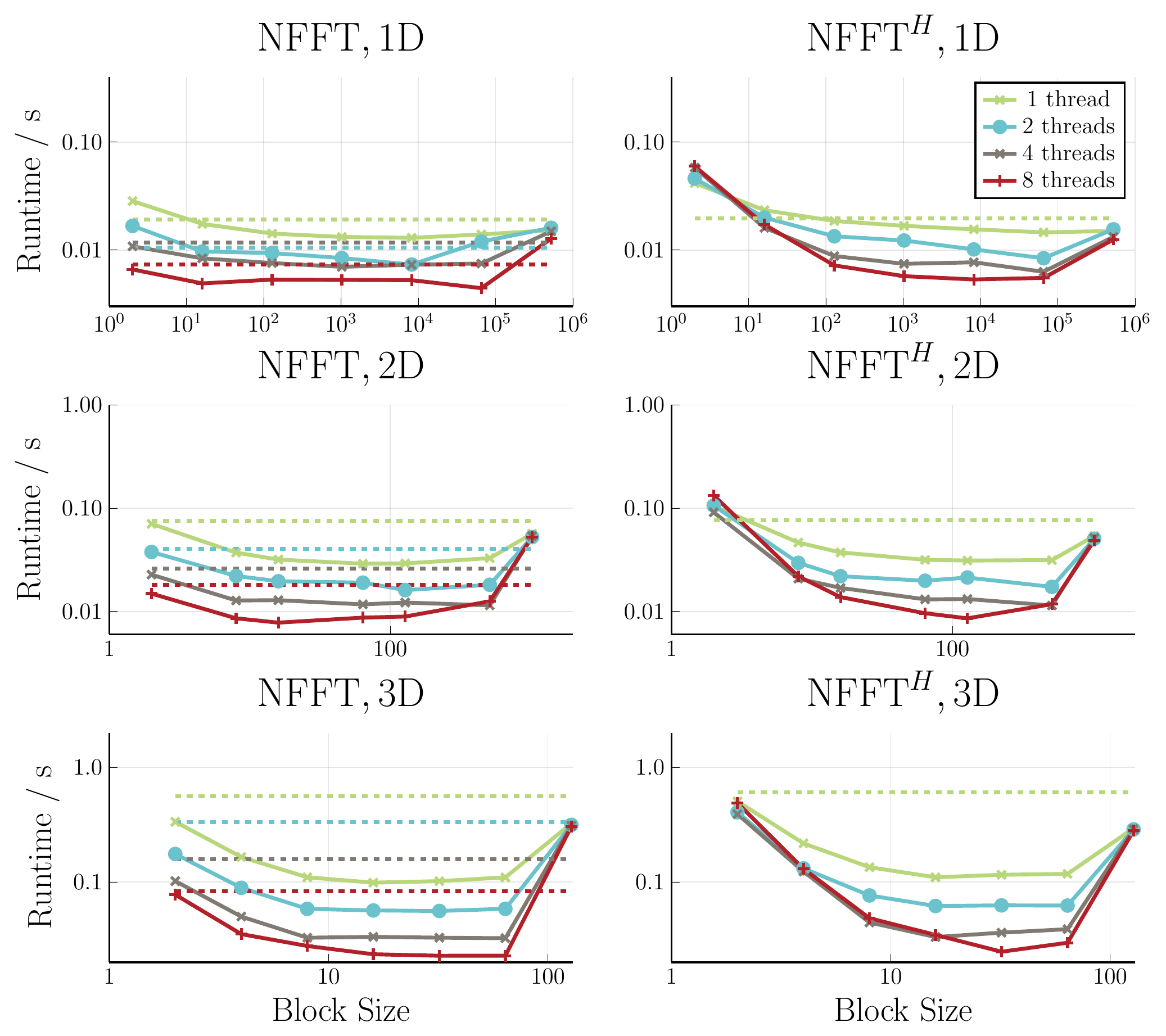}
\vspace{-0.7cm}
  \caption{Runtime performance of  NFFT.jl for  block-partitioned (solid) and regular (dashed) resampling, 1D--3D transformations, and 1--8 threads. Left shows the runtime of the direct transformation while right shows the runtime of the adjoint transformation. }
  \label{fig:performanceBlockSize}
\end{figure}

We next take a closer look at the resampling and investigate the block-partitioning performance within NFFT.jl. \Cref{fig:performanceBlockSize} shows the runtime performance of the direct and the adjoint NFFT for different block sizes, different dimensionalities (1D--3D), \texttt{POLYNOMIAL} precomputation, and different numbers of threads (1--8). The block sizes are chosen in such a way, that the largest value (1D: $2^{19}$, 2D: $(1024,1024)^\intercal$, 3D: $(128, 128, 128)^\intercal$) corresponds to the usage of one block only. This edge case results in no parallelization and thus leads to sub-optimal performance in the multi-threaded cases. When choosing smaller block sizes also the single-threaded performance is improved for all dimensionalities since this allows for better usage of CPU caches. For all dimensionalities, one can see that the performance degrades when choosing the blocks too small since in that case, the administration overhead for the blocks becomes too high. The optimum block size differs for different numbers of threads and the direct and adjoint transform. It ranges from $10^3$ to $10^5$ in 1D, from $64$ to $128$ in 2D, and from $16$ to $32$ in 3D. Based on that we set the default value to $4096$ (1D), $(64,64)^\intercal$ (2D), and $(16, 16, 16)^\intercal$ (3D) in NFFT.jl. For higher dimensions the block size is set to one.

In addition to the block-partitioning performance (solid), the figure also shows the performance of the regular resampling (dashed). For the adjoint transform the latter is only available for single-threading. One can see that the block-partitioned resampling outperforms the regular resampling in all cases when considering an optimal block size. This shows that block-partitioning should not only be used for the adjoint but also for the direct NFFT.

\subsection{Performance Analysis}
\begin{figure}[t!]
  \centering
  \includegraphics[width=\textwidth]{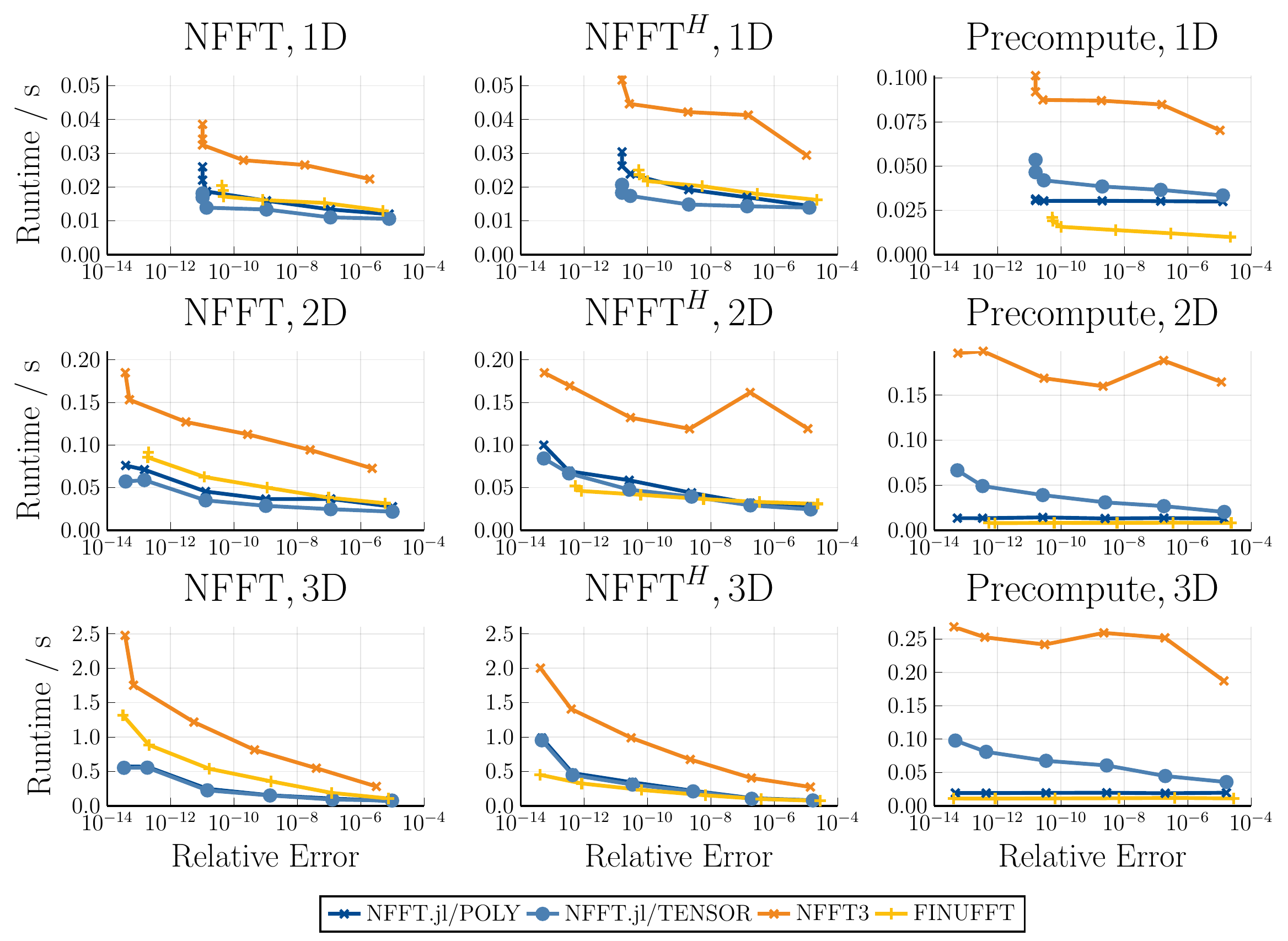}
  \vspace{-0.7cm}
  \caption{Single-threading performance of NFFT.jl with the TENSOR and the POLYNOMIAL precomputation strategy compared to NFFT3 (with TENSOR) and FINUFFT.}
  \label{fig:performanceSingleThread}
\end{figure}

A performance comparison of NFFT.jl, NFFT3, and FINUFFT for 1D--3D and running on a single thread is shown in  \cref{fig:performanceSingleThread}. In addition to the runtime performance (columns 1--2) also the precomputation time (column 3) is shown. All times are calculated for $m=3,\dots, 8$ and plotted versus the relative error. We use this representation, since it allows us to make comparisons even if the accuracy of different implementations differs.

The runtime of NFFT.jl is shown for both \texttt{POLYNOMIAL} and \texttt{TENSOR} pre-com\-pu\-ta\-tion. One can see that the runtime performance is usually a little bit higher for \texttt{TENSOR} with the downside of a larger precomputation time. This trade-off is more prominent for the 1D and 2D transform whereas in 3D the runtimes are very close while the precomputation times are almost negligibly small compared to the transformation time. As a result, we  decided to make \texttt{POLYNOMIAL} the default option since it provides good performance and uses less memory. 

When comparing the different libraries one can see that the performance varies for different dimensionality and different accuracy. In all cases, FINUFFT and NFFT.jl are faster than NFFT3. In 1D, NFFT.jl with  \texttt{POLYNOMIAL} precomputation is as fast as FINUFFT while  \texttt{TENSOR} precomputation is fastest. In 2D, \texttt{TENSOR} precomputation is still a little bit faster. In that case, FINUFFT is slighly slower in the direct transform while for the adjoint transform NFFT.jl and FINUFFT are almost on the same level with FINUFFT being slightly faster for high accuracy. The results in 3D are similar, with the only exception that the NFFT.jl with  \texttt{POLYNOMIAL} precomputation is at the same level as NFFT.jl  with \texttt{TENSOR} precomputation. This is expected because the local window caching discussed in section~\ref{sec:loopoptim} exploits the tensor product structure, resulting in a lower time fraction for the window precomputation than for the actual summation in the resampling step.


When looking at the precomputation time one can see that FINUFFT is fastest for all dimensionalities. This is most important for 1D transforms where the precomputation time is larger or in the same order as the actual transformation time. 



\begin{figure}[t!]
  \centering
  \includegraphics[width=\textwidth]{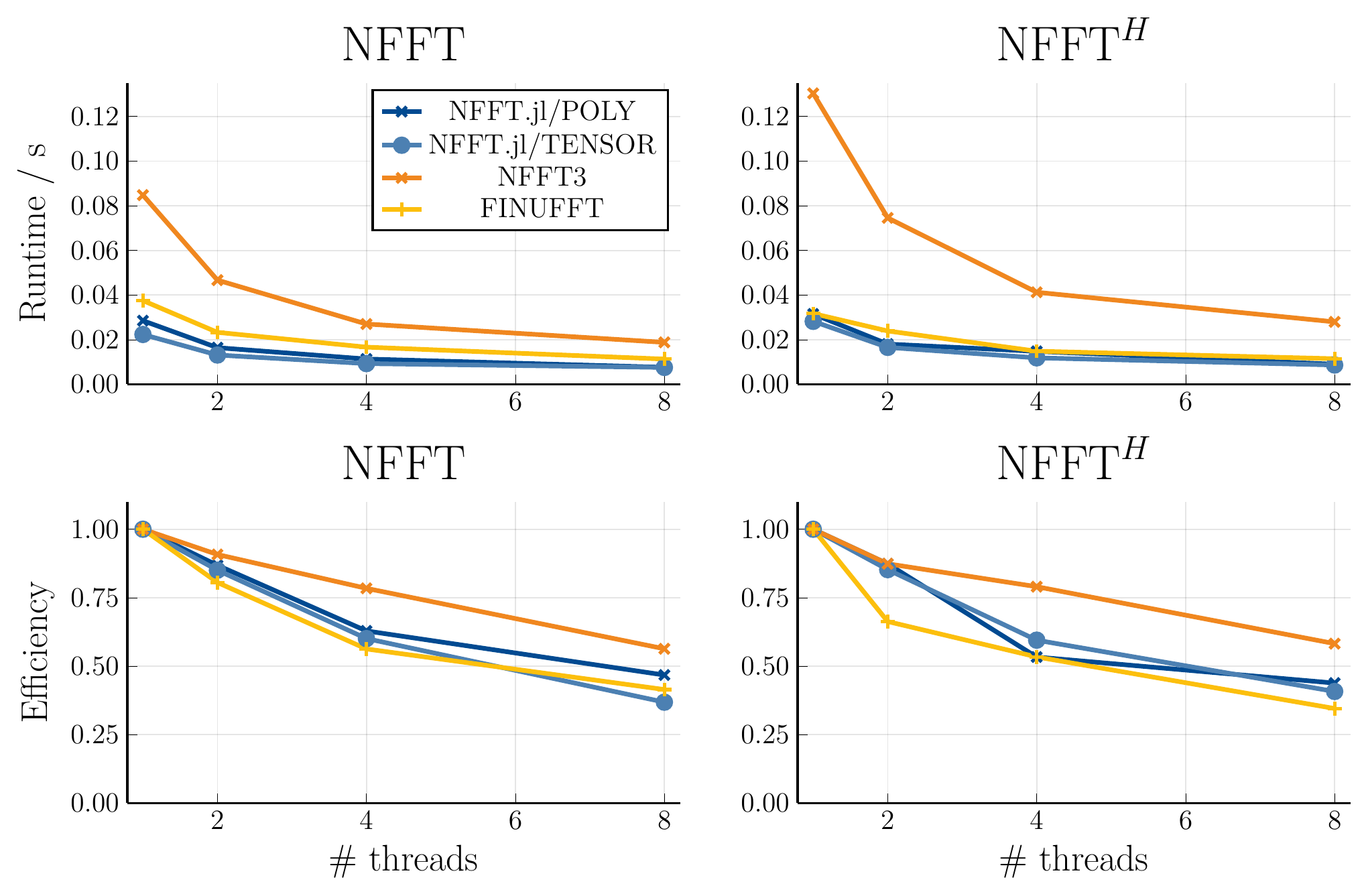}
  \vspace{-0.7cm}
  \caption{Multi-threading performance of NFFT.jl with the TENSOR and the POLYNOMIAL precomputation strategy compared to NFFT3 (with TENSOR) and FINUFFT. Shown is the speedup and the parallel efficiency for $t=1,2,4,8$ threads. }
  \label{fig:performanceMultiThread}
\end{figure}

The multi-threading performance of NFFT.jl, NFFT3, and FINUFFT is compared in \cref{fig:performanceMultiThread} for the 2D data ($m=4$ and $\sigma=2$). The upper two plots show the runtime for $t=1,\dots, 8$ threads. One can see that all libraries speedup computation by adding more threads although the parallel efficiency (lower two plots) drops with increasing threads. The parallel efficiency is defined as the ratio between actual and theoretically possible speedup. It is in a similar range for all three libraries with a slightly higher value for NFFT3. This might be due to the lock required in NFFT.jl's and FINUFFT's block-partitioning implementation, which is not required in NFFT3's implementation.

\section{Discussion} \label{Sec:Discussion}

The aim of this work was to introduce the software package NFFT.jl, which is written in pure Julia and combines high-performance with a flexible architecture. The use of Julia is not an implementation detail but, on the contrary, enables us to make less compromises in the implementation, which would neither be possible in high-level languages like Matlab and Python, nor in low-level languages like C/C++. Along these lines, our work can also be understood as a showcase of how Julia enables new possibilities in scientific computing that are challenging to achieve with classical tools. The highlight of NFFT.jl is a completely generic implementation that is both number-type and dimension-agnostic.

A comparison of code complexity is not really possible since it also depends on the programmers experience in a certain programming language, whether a code base is found to be complex. Even a comparison of code size is difficult, since the number of characters also depends on the verbosity of keywords used in a given language. Keeping that in mind we counted memory used by the source code of the core NFFT algorithm for all three libraries. It is 184 KB for the \texttt{kernel/nfft} directory of NFFT3 and 246 KB for the \texttt{src} directory of FINUFFT. The latter, however, contains 106 KB of generated code, which we would not count as source code yielding 140 KB effectively. In contrast, the \texttt{src} directory of NFFT.jl has 74 KB of source code. This shows that Julia allows for using less code for achieving the same or even more functionality than its C/C++ counterparts.

The performance of NFFT.jl was found to be the same or better than NFFT3 and FINUFFT. NFFT.jl was faster in both the 1D and the 2D examples and was at the same speed as FINUFFT for the 3D example. Regarding precomputation time, NFFT.jl was only slightly slower than FINUFFT and much faster than NFFT3 for the \texttt{POLYNOMIAL} precomputation. NFFT.jl with \texttt{TENSOR} precomputation was slower than \texttt{POLYNOMIAL} but still much faster than NFFT3, which is due to our polynomial window approximation being applied during precomputation.

Beside the benchmark between different NFFT software libraries we also made some general findings on NFFT implementation strategies. First of all we compared the use of $2m$ and $2m+2$ sampling points for the window function and found that the gain in accuracy by the two additional points sampled outside the support of $\hat{\psi}$ are not worth the additional accuracy, since it is more effective to instead increase the window parameter $m$ itself, which increases the width of the window. Moreover, we investigated different block sizes and found dimension-dependent value ranges for which the NFFT reaches highest performance. Since the optimum block-size is highly dependent on the transform size, the number of used threads, and the CPU, we plan to develop an optimization mode, where the optimum block size is automatically chosen based on online-benchmarks, similar to the \texttt{FFTW\_MEASURE} option in FFTW. 




One principle downside of NFFT.jl compared to its C/C++ counterparts is that the latter are currently more binding-friendly than the former. We note, however, that Julia can be embedded in C and therefore it is possible to integrate NFFT.jl in every programming language that allows to call C code. While this still requires a full Julia installation there is also the possibility to statically compile Julia code into a shared library using PackageCompiler.jl. We note, however, that static compilation of Julia code is still actively being worked on and will likely evolve in future versions of the language.

An important future step for the NFFT.jl project is to exploit GPU implementations written in Julia. Right now, there is a prototype implementation named CuNFFT.jl, which is fully functional and required only 125 lines of code. This is achieved by using the \texttt{FULL} precomputation operation and uploading the entire sparse matrix onto the GPU. For optimum performance, one however, would need to write custom kernels and use a similar block-partitioning strategy as we did in the CPU implementation.

NFFT.jl is already a feature-rich NFFT library that is used in  different contexts \cite{schutz2015large,jacobsen2016generalized,knopp2021mrireco}. During the write-up of this paper, we focused on fixing performance bottlenecks and streamlining the interface of the core algorithm. This does not imply, that NFFT.jl is feature complete yet. One important extension would be the implementation of the NNFFT (also named type-3 NFFT), which has non-equidistant sampling points in both domains. Another missing feature are the fast versions of the non-equidistant cosine and sine transforms (NFCT and NFST). Application oriented tools -- like quadrature weights for the non-equidistant sampling points -- are also in the scope of NFFT.jl. Those are implemented in the package NFFTTools.jl, which is one layer lower in the package graph and can in principle be also used in combination with other NFFT Julia packages such as NFFT3.jl and FINUFFT.jl.




\section{Conclusion}

In conclusion, this work has outlined how the scientific programming language Julia can be used to implement a very flexible software package for NFFT computation. Our implementation is completely type- and dimension-agnostic and still uses less code than established packages. We implemented state-of-the-art acceleration techniques taken from two established NFFT libraries and showcased their performance characteristic in 1D--3D examples.

\section*{Acknowledgments}
We would like to thank all developers and users who have contributed to NFFT.jl. Furthermore, we thank Daniel Potts, Alexander Barnett, Toni Volkmer, Michael Quellmalz and Michael Schmischke for discussions and hints on the use of NFFT3 and FINUFFT. Finally, we thank Jeffrey A. Fessler for fruitful discussions on NFFT terminology.

\bibliographystyle{siamplain}

\begin{thebibliography}{10}

\bibitem{anderson1996rapid}
{\sc C.~Anderson and M.~D. Dahleh}, {\em Rapid computation of the discrete
  {Fourier} transform}, SIAM Journal on Scientific Computing, 17 (1996),
  pp.~913--919.

\bibitem{barnett2019parallel}
{\sc A.~H. Barnett, J.~Magland, and L.~af~Klinteberg}, {\em A parallel
  nonuniform fast {Fourier} transform library based on an ``exponential of
  semicircle'' kernel}, SIAM Journal on Scientific Computing, 41 (2019),
  pp.~C479--C504.

\bibitem{beylkin1995fast}
{\sc G.~Beylkin}, {\em On the fast {Fourier} transform of functions with
  singularities}, Applied and Computational Harmonic Analysis, 2 (1995),
  pp.~363--381.

\bibitem{bezanson2017julia}
{\sc J.~Bezanson, A.~Edelman, S.~Karpinski, and V.~B. Shah}, {\em Julia: A
  fresh approach to numerical computing}, SIAM review, 59 (2017), pp.~65--98.

\bibitem{cooley1965algorithm}
{\sc J.~W. Cooley and J.~W. Tukey}, {\em An algorithm for the machine
  calculation of complex {Fourier} series}, Mathematics of computation, 19
  (1965), pp.~297--301.

\bibitem{dutt1993fast}
{\sc A.~Dutt and V.~Rokhlin}, {\em Fast {Fourier} transforms for nonequispaced
  data}, SIAM Journal on Scientific computing, 14 (1993), pp.~1368--1393.

\bibitem{fessler2007nufft}
{\sc J.~A. Fessler}, {\em On nufft-based gridding for non-cartesian {MRI}},
  Journal of magnetic resonance, 188 (2007), pp.~191--195.

\bibitem{fessler2003nonuniform}
{\sc J.~A. Fessler and B.~P. Sutton}, {\em Nonuniform fast fourier transforms
  using min-max interpolation}, IEEE transactions on signal processing, 51
  (2003), pp.~560--574.

\bibitem{FFTW05}
{\sc M.~Frigo and S.~G. Johnson}, {\em The design and implementation of
  {FFTW3}}, Proceedings of the IEEE, 93 (2005), pp.~216--231.
\newblock Special issue on ``Program Generation, Optimization, and Platform
  Adaptation''.

\bibitem{hillmann2009using}
{\sc D.~Hillmann, G.~Huttmann, and P.~Koch}, {\em Using nonequispaced fast
  {Fourier} transformation to process optical coherence tomography signals}, in
  European Conference on Biomedical Optics, Optical Society of America, 2009,
  p.~7372\_0R.

\bibitem{jacobsen2016generalized}
{\sc R.~D. Jacobsen, M.~Nielsen, and M.~G. Rasmussen}, {\em Generalized
  sampling in julia}, arXiv preprint arXiv:1607.04091,  (2016).

\bibitem{johnson2009convolution}
{\sc K.~O. Johnson and J.~G. Pipe}, {\em Convolution kernel design and
  efficient algorithm for sampling density correction}, Magnetic Resonance in
  Medicine: An Official Journal of the International Society for Magnetic
  Resonance in Medicine, 61 (2009), pp.~439--447.

\bibitem{keiner2009using}
{\sc J.~Keiner, S.~Kunis, and D.~Potts}, {\em Using {NFFT} 3---a software
  library for various nonequispaced fast fourier transforms}, ACM Transactions
  on Mathematical Software (TOMS), 36 (2009), pp.~1--30.

\bibitem{knoll2014gpunufft}
{\sc F.~Knoll, A.~Schwarzl, C.~Diwoky, and D.~K. Sodickson}, {\em {gpuNUFFT} -
  an open source {GPU} library for {3D} regridding with direct {Matlab}
  interface}, in International Society for Magnetic Resonance in Medicine:
  Scientific Meeting \& Exhibition, 2014, pp.~4297--4297.

\bibitem{knopp2014experimental}
{\sc T.~Knopp}, {\em Experimental multi-threading support for the julia
  programming language}, in Proceedings of the First Workshop for High
  Performance Technical Computing in Dynamic Languages, IEEE Press, 2014,
  pp.~1--5.

\bibitem{knopp2021mrireco}
{\sc T.~Knopp and M.~Grosser}, {\em {MRIReco.jl}: An {MRI} reconstruction
  framework written in julia}, Magnetic resonance in medicine, 86 (2021),
  pp.~1633--1646.

\bibitem{knopp2007note}
{\sc T.~Knopp, S.~Kunis, and D.~Potts}, {\em A note on the iterative {MRI}
  reconstruction from nonuniform k-space data}, International journal of
  biomedical imaging, 2007 (2007).

\bibitem{kunis2012nonequispaced}
{\sc S.~Kunis and S.~Kunis}, {\em The nonequispaced {FFT} on graphics
  processing units}, Pamm, 12 (2012), pp.~7--10.

\bibitem{kunis2008time}
{\sc S.~Kunis and D.~Potts}, {\em Time and memory requirements of the
  nonequispaced fft}, Sampling Theory in Signal and Image Processing, 7 (2008),
  pp.~77--100.

\bibitem{potts2001fast}
{\sc D.~Potts, G.~Steidl, and M.~Tasche}, {\em Fast {Fourier} transforms for
  nonequispaced data: A tutorial}, Modern sampling theory,  (2001),
  pp.~247--270.

\bibitem{potts2021continuous}
{\sc D.~Potts and M.~Tasche}, {\em Continuous window functions for {NFFT}},
  Advances in Computational Mathematics, 47 (2021), pp.~1--34.

\bibitem{potts2021uniform}
{\sc D.~Potts and M.~Tasche}, {\em Uniform error estimates for nonequispaced
  fast {Fourier} transforms}, Sampling Theory, Signal Processing, and Data
  Analysis, 19 (2021), pp.~1--42.

\bibitem{schutz2015large}
{\sc A.~Schutz, A.~Ferrari, D.~Mary, {\'E}.~Thi{\'e}baut, and F.~Soulez}, {\em
  Large scale {3D} image reconstruction in optical interferometry}, in 2015
  23rd European Signal Processing Conference (EUSIPCO), IEEE, 2015,
  pp.~474--478.

\bibitem{shih2021cufinufft}
{\sc Y.-h. Shih, G.~Wright, J.~And{\'e}n, J.~Blaschke, and A.~H. Barnett}, {\em
  {cuFINUFFT}: a load-balanced {GPU} library for general-purpose nonuniform
  {FFTs}}, in 2021 IEEE International Parallel and Distributed Processing
  Symposium Workshops (IPDPSW), IEEE, 2021, pp.~688--697.

\bibitem{sorensen2008accelerating}
{\sc T.~S. S{\o}rensen, T.~Schaeffter, K.~{\O}. Noe, and M.~S. Hansen}, {\em
  Accelerating the nonequispaced fast {Fourier} transform on commodity graphics
  hardware}, IEEE Transactions on Medical Imaging, 27 (2008), pp.~538--547.

\bibitem{steidl1998note}
{\sc G.~Steidl}, {\em A note on fast {Fourier} transforms for nonequispaced
  grids}, Advances in computational mathematics, 9 (1998), pp.~337--352.

\bibitem{volkmer2012openmp}
{\sc T.~Volkmer}, {\em {OpenMP} parallelization in the {NFFT} software
  library}, Preprint,  (2012),
  \url{https://www-user.tu-chemnitz.de/~potts/paper/openmpNFFT.pdf}.

\bibitem{ware1998fast}
{\sc A.~F. Ware}, {\em Fast approximate {Fourier} transforms for irregularly
  spaced data}, SIAM review, 40 (1998), pp.~838--856.

\bibitem{yang2018new}
{\sc S.-C. Yang, H.-J. Qian, and Z.-Y. Lu}, {\em A new theoretical derivation
  of {NFFT} and its implementation on {GPU}}, Applied and Computational
  Harmonic Analysis, 44 (2018), pp.~273--293.

\end{thebibliography}

\end{document}